\documentclass[journal]{IEEEtran}
\ifCLASSINFOpdf
\else
\fi
\usepackage{breakcites}
\usepackage{notoccite}
\usepackage{multirow}
\usepackage{url} 
\usepackage{float}
\usepackage{wrapfig} 
\usepackage{graphicx}
\usepackage[noadjust]{cite}
\usepackage[cmex10]{amsmath}
\usepackage{enumitem}
\usepackage{comment}
\usepackage[cmintegrals]{newtxmath}
\usepackage{algorithm}
\usepackage[noend]{algpseudocode}
\usepackage[caption=false]{subfig}
\vbadness=\maxdimen

\begin{document}
\title{An iBeacon based Proximity and Indoor Localization System}
\author{Faheem~Zafari,~\IEEEmembership{Student~Member,~IEEE,}
        Ioannis~Papapanagiotou,~\IEEEmembership{Senior~Member,~IEEE,}
        Michael~Devetsikiotis,~\IEEEmembership{Fellow,~IEEE,}
        and~Thomas~Hacker,~\IEEEmembership{Member,~IEEE}
\thanks{Faheem Zafari is with Department of Electrical and Electronics Engineering, Imperial College, London,
UK e-mail: faheem16@imperial.ac.uk}
\thanks{Ioannis Papanagiotou is with Netflix, Los Gatos, CA, USA email: ipapapa@ncsu.edu}
\thanks{Michael Devetsikiotis is with Department of Electrical and Computer Engineering, University of New Mexico, Albuquerque, NM, USA email: mdevets@unm.edu}
\thanks{Thomas Hacker is with Department of Computer and Information Technology, Purdue University, West Lafayette, IN, USA email: tjhacker@purdue.edu}}

\markboth{}%
{Shell \MakeLowercase{\textit{et al.}}: Bare Demo of IEEEtran.cls for IEEE Journals}

\maketitle

\begin{abstract}

Indoor localization and \textit{Location Based Services} (LBS) can greatly benefit from the widescale proliferation of communication devices. The basic requirements of a system that can provide the aforementioned services are energy efficiency, scalability, lower costs, wide reception range, high localization accuracy and availability. Different technologies such as WiFi, UWB, RFID have been leveraged to provide LBS and \textit{Proximity Based Services} (PBS), however they do not meet the aforementioned requirements. Apple's Bluetooth Low Energy (BLE) based \textit{iBeacon} solution primarily intends to provide \textit{Proximity Based Services} (PBS). However, it suffers from poor proximity detection accuracy due to its reliance on Received Signal Strength Indicator (RSSI) that is prone to multipath fading and drastic fluctuations in the indoor environment. 
Therefore, in this paper, we present our iBeacon based accurate proximity and indoor localization system. Our two algorithms \textit{Server-Side Running Average} (SRA) and \textit{Server-Side Kalman Filter} (SKF) improve the proximity detection accuracy of iBeacons by 29\% and 32\% respectively, when compared with Apple's current moving average based approach. We also present our novel cascaded Kalman Filter-Particle Filter (KFPF) algorithm for indoor localization.  Our cascaded filter approach uses a Kalman Filter (KF) to reduce the RSSI fluctuation and then inputs the filtered RSSI values into a Particle Filter (PF) to improve the accuracy of  indoor localization. Our experimental results, obtained through experiments in a space replicating real-world scenario, show that our cascaded filter approach outperforms the use of only PF by 28.16\%  and 25.59\% in 2-Dimensional (2D) and 3-Dimensional (3D) environments respectively, and achieves a localization error as low as 0.70 meters in 2D environment and 0.947 meters in 3D environment. 

\end{abstract}


\IEEEpeerreviewmaketitle

\section{Introduction}
The rapid developments in the field of communication and networking has resulted in a wide range of different services with the intent to improve the overall Quality of Service (QoS) provided to the users \cite{faheem2015IoT}. Location and proximity based services (PBS) are examples of such services that have recently witnessed an increase in interest, particularly after the advent of the Internet of Things (IoT). The motive behind Location Based Services (LBS) and PBS is to leverage the user location or proximity to provide targeted and relevant services such as automating various devices or appliances based on the user location.  While outdoor localization has been extensively researched and there is already a widely accepted outdoor localization system called the \textit{Global Positioning System} (GPS), indoor localization is a relatively novel field of research that currently lacks a widely adopted standardized system. 
\par Indeed, indoor localization is much more challenging because of the presence of wide range of obstacles including walls, people etc. that results in increased \textit{multipath} signals and effects. Furthermore, the localization or proximity detection accuracy requirement for indoor environments is below one meter while for GPS, it is about 5-10 meters \cite{faheem2015IoT}. This is because 5-10 meters accuracy is feasible for street level navigation, however, for indoor environments such as a meter wide isle of library, we cannot tolerate a large localization/proximity error \cite{xiong2013arraytrack}.  Such stringent accuracy requirements, along with lower cost, higher reception range, availability, energy efficiency and scalability makes indoor localization and proximity detection a challenging research problem. Solutions that rely on WiFi, Radio Frequency Identification (RFID), Ultra-wide band (UWB), ultrasound, BLE etc. have been proposed in the literature. However, they do not fulfil the aforementioned requirements. Particularly, since localization and proximity rely heavily on the user device, the energy consumption on the user device is of fundamental importance. Therefore, there is a need for an accurate, cost and energy efficient localization and proximity detection system that can be leveraged to provide LBS and PBS. Apple's BLE based \textit{iBeacon} protocol is primarily designed to provide PBS. A proprietary application on the user device listens to the messages transmitted by the \textit{iBeacons} (the device which runs the iBeacon protocol is called beacon/iBeacon) and then uses a moving average of the Received Signal Strength Indicator (RSSI) to estimate the proximity of the user to any specific beacon. RSSI is the most affordable and widely used metric to obtain an estimate of the distance between a user  and the beacon as it does not require complex calculation. However, it is prone to the multipath effects and noise which significantly reduces its localization/proximity detection accuracy. \par
In this paper, we discuss our iBeacon based proximity and indoor localization system. We present two algorithms \textit{Server-side Running Average} (SRA) and \textit{Server-side Kalman Filter (SKF)} that mitigate the inherent problems of iBeacon based proximity detection and improves its accuracy. Our proximity based system is an extended version of our work in \cite{faheemsensys}\footnote{The paper is titled ``Enhancing the accuracy of iBeacons for Indoor Proximity-based services'' and  will be presented in IEEE International Conference on Communications (ICC) 2017} and builds on our prior work on particle filters \cite{faheemglobecom}.  We also use iBeacons for indoor localization, despite the fact that they are not designed for indoor localization.  We leverage Bayesian filtering based \textit{Particle Filter} (PF) and \textit{Kalman Filter} (KF) to improve the performance of an iBeacon based indoor localization system. Experimental results show that our proposed approach of using cascaded KF and PF outperforms using only a PF for 2D and 3D indoor localization.  The main contributions of this paper are:
\begin{itemize}
	\item Utilizing iBeacons for accurate indoor localization despite the fact that they are primarily intended for proximity detection.
	\item Designing and implementing two algorithms, SRA and SKF, that improve the proximity detection accuracy of iBeacons by 29\% and 32\% respectively when compared with the current moving averaged based approach used by iBeacons. 
	\item Designing and implementing KFPF cascaded algorithm to improve the localization accuracy by 28.16\% and 25.59\% when compared with using only a PF in 2D and 3D environments respectively.
\end{itemize}
The paper is further structured as: Section \ref{sec:back} presents a review of the literature along with a primer on iBeacons. Section \ref{3} discusses Bayesian filtering, indoor localization, KF and PF. Section \ref{sec:proxandlocalsys} presents our iBeacon based proximity and indoor localization system. Section \ref{sec:expres} presents our experimental results.
 Section \ref{sec:conclusion} concludes the paper. 

\section{Background and Related Work}
\label{sec:back}
In this section, we first discuss some of the localization and proximity solutions proposed in the literature. We also present a primer on \textit{iBeacons} and describe Bayesian Filtering in detail. 
\subsection{Related Work}
Indoor localization and proximity detection has recently witnessed an increase in interest. A number of solutions \cite{vasisht2016decimeter,xiong2013arraytrack,xiong2015tonetrack,kumar2014accurate,faheemglobecom} have been proposed in the literature that rely on a number of technologies including WiFi, UWB, RFID, and BLE etc. and metrics such as RSSI, Angle of Arrival (AoA), Time of Flight (ToF), Time Difference of Arrival (TDoA) etc. for accurate indoor localization and tracking. However, they do not satisfy the aforementioned requirements required for localization. \par
Kumar et al. propose Ubicarse \cite{kumar2014accurate} that leverages existing WiFi Access Points (AP) for accurately localizing a user device using AoA. Ubicarse emulates Synthetic Aperture Radar (SAR) on a user device through a novel formulation that is resilient to translation motion of the device. By rotating the device, the user is able to obtain his location with respect to different WiFi APs. Ubicarse attains median localization accuracy of 39 cm and also provides proximity based services. However, Ubicarse requires two antennas on the user device. The user has to rotate the device to emulate SAR, which is not ideal for real world applications. 
Xiong et al. propose an AoA based system called ArrayTrack \cite{xiong2013arraytrack}  that uses a modified version of the the widely used MUSIC algorithm \cite{schmidt1986multiple} to obtain a relatively accurate estimate of the AoA spectrum. To further refine the performance of the system, ArrayTrack leverages \textit{spatial smoothing}. ArrayTrack requires only a small number of packets to track the user in real time (100ms) which to the best of our knowledge is currently the only system that can provide accurate localization in real time. ArrayTrack achieves a median localization accuracy of 23cm. ArrayTrack as of now is based on proprietary hardware and has not been tested with off the shelf WiFi cards. Therefore, it currently incurs added cost. ArrayTrack also requires higher number of antennas that currently might not be supported by majority of WiFi APs. Xiong et al. also propose ToneTrack \cite{xiong2015tonetrack} that uses ToA and TDoA information to provide fairly accurate real time indoor localization. ToneTrack combines different WiFi channels as the user device hops frequencies to obtain a fine time resolution information required for indoor localization. ToneTrack attains a median accuracy of 0.9m, however it currently relies on proprietary hardware to emulate a WiFi AP. Existing WiFi APs are yet to be tested with ToneTrack.  Vashist et al. \cite{vasisht2016decimeter} propose a ToF based localization system called \textit{Chronos} that uses a single WiFi AP to localize a user. Chronos, like ToneTrack, combines the information across different WiFi channels to obtain fine time resolution like UWB. Chronos achieves a median localization accuracy of 65 cm. While Chronos has been used for proximity based services, it is yet to be seen how scalable it is and whether it can effectively be used in real time. In our prior work \cite{faheemglobecom}, we used only particle filtering on the user device to provide indoor localization using iBeacons. An accuracy as high as 97cm was obtained. However, using PF on the user device significantly drains the battery and challenges the processing power of the user device. An earlier version of this work \cite{faheemsensys} only provided an iBeacon based proximity detection system. 
\par Unlike Ubicarse \cite{kumar2014accurate}, our proposed iBeacon based proximity and indoor localization system does not require the user to twist his device for localization. Furthermore, by offloading the algorithms to the server, our proposed approach consumes lesser energy than Ubicarse. In comparison with ArrayTrack \cite{xiong2013arraytrack}, and ToneTrack \cite{xiong2015tonetrack}, our proposed system does not require expensive hardware or added number of antennas. We used iBeacons which cost as low as \$5 (Gimbal Series 10 \cite{gimbal}). In contrast with Chronos \cite{vasisht2016decimeter}, our proposed system does not affect the performance of existing WiFi architecture and is scalable.  While our prior work \cite{faheemglobecom} used only PF on the user device, this paper uses a server side cascaded filtering approach that improves the localization accuracy. By offloading the algorithms to a server, we reduce the energy consumption on the user device and leverage the higher processing power of the server to improve the localization latency. Furthermore, this work also provides 3D localization and proximity detection.

\subsection{A Primer on iBeacons}
\begin{figure}
 	\centering
 	\includegraphics[width=0.48\textwidth]{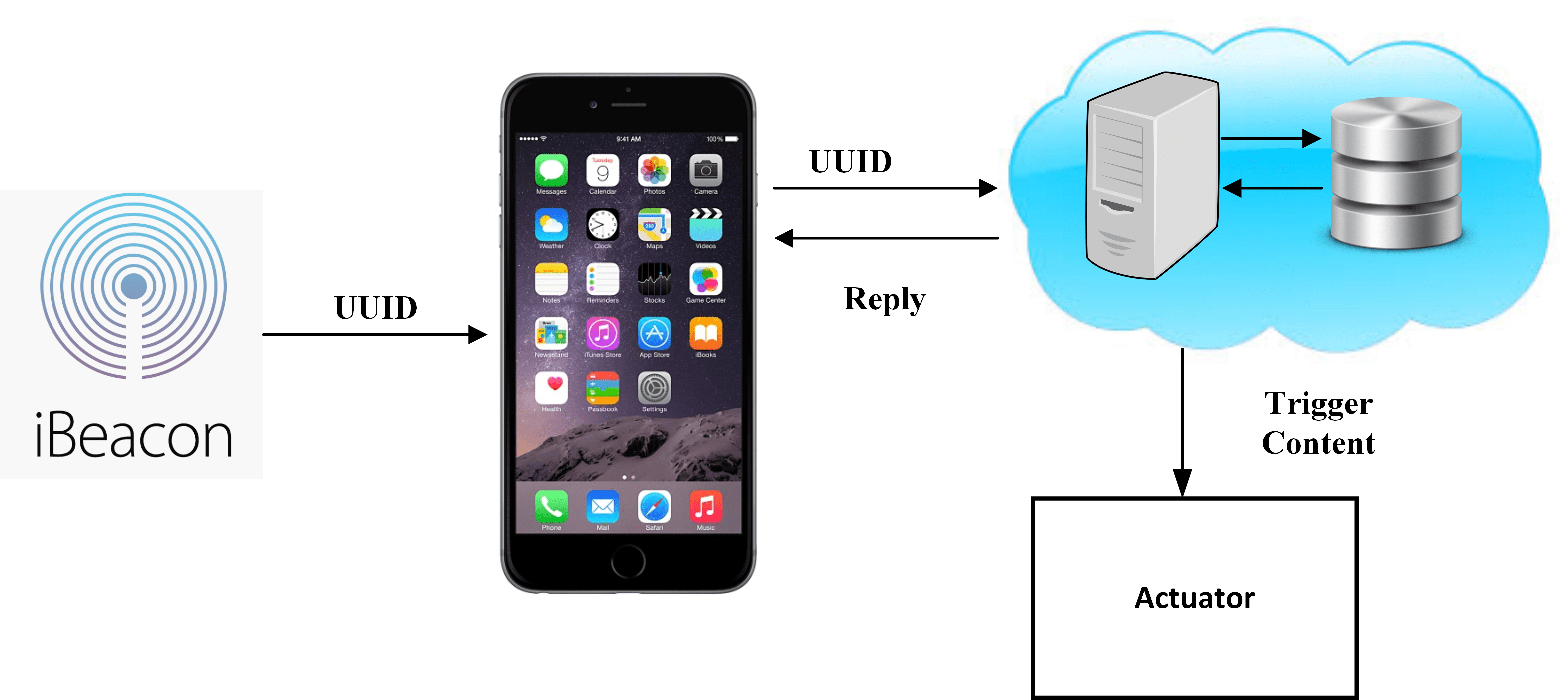}
 	\caption{Working principle of the iBeacon}
 	\protect\label{fig:1}
 	\vspace{-20pt}
 \end{figure}

 iBeacons/Beacons are small and energy efficient BLE enabled devices that use Apple's proprietary iBeacon protocol \cite{appleibeaconspecification}. The iBeacon protocol allows any BLE enabled device (such as a smartphone) to receive a signal from the beacons. The message transmitted by iBeacons consists of 
 \begin{itemize}
 	\setlength\itemsep{0em}
 	\item A Universally Unique Identifier (UUID): The UUID string  helps to identify the beacons used by any particular company. (Mandatory) 
 	\item A Major Value: The next part is a major value that helps to differentiate beacons of a specific brand `X' present in a location such as a city `Y'. (Optional)   
 	\item A Minor Value: The minor value helps to identify the beacon of any brand `X', in city `Y' and department `Z'. (Optional)
 \end{itemize}  
 User devices that are BLE enabled and running either iOS 7.0+ or Android 4.3+ operating systems can be used for beacon related services \cite{appleibeaconspecification}. Specific mobile applications that are beacon capable can be developed to communicate with the beacons.
 There is no limit to the number of devices that can be present in a space. However, a single device can communicate with more than 4 billion iBeacons. Using RSSI, the user's proximity to the beacon is classified in one of the four zones listed in Table \ref{tab:proximityzones}.
 Once the user device obtains a UUID, it contacts a server to inquire about the UUID and the event associated with the iBeacon. The server responds back with relevant information and can trigger an event such as responding back to the user with a discount coupon or opening a security door based on the user's proximity to the door. Figure \ref{fig:1} shows the working principle of the iBeacon. 
 \par As mentioned earlier, Apple's current proximity detection mechanism is based on the CoreLocation framework for iOS application development \cite{appleibeaconspecification}. 
 Different applications could potentially use any of the aforementioned zones to provide PBS. It is therefore fundamental to accurately compute the user's proximity to the iBeacons.  A user who enters a store such as `Starbucks'  and is in the `immediate` zone of the counter, could avoid lengthy queues by leveraging his accurate proximity to the iBeacons. The user can confirm the order through his smartphone and pay for the order based on his proximity. Similarly, a user could book a hotel room online and then confirm his arrival through the iBeacon at the main entrance of the hotel without going through any other check-in procedure at the hotel. The elevator would be called up based on the user's location and he would be taken to his floor. When the user is within the `immediate' vicinity of his room, the door would automatically open up eliminating the need for using any card or key. Such services are only possible with accurate proximity estimation and the proximity error being within certain bounds. However, using the current approach, the RSSI values from the beacons are highly fluctuating that increases the probability of misclassification. Therefore, there is a need for filtering mechanism that can stabilize the RSSI values and improve the overall performance of iBeacons. 
\begin{table}[]
	\centering
	\caption{The classification of proximity zones based on distance between the user and the iBeacon}
	\label{tab:proximityzones}
	\begin{tabular}{|l|l|}
		\hline
		\textbf{Zone} & \textbf{Distance} \\ \hline
		Immediate     & \textless 1 m      \\ \hline
		Near          & 1-3 m             \\ \hline
		Far           & \textgreater3 m   \\ \hline
		Unknown       & Device not ranged \\ \hline
	\end{tabular}
\vspace{-12pt}
\end{table}

\section{Bayesian Filters and Indoor Localization\label{3}}
A Bayesian filtering approach can be used for indoor localization, because the user position (state) varies dynamically with time resulting in a dynamic system problem \cite{arulampalam2002tutorial}. The two models required for dynamic system problems are: 
\begin{itemize}
	\item \textit{System Model}: The system model relates the evolution of state to time. In terms of indoor localization, it signifies how the user location changes with time. In terms of proximity, it signifies how the RSSI values change with time. 
	\item \textit{Measurement Model}: The measurement model relates the obtained noisy measurements, obtained from sensors, with the state. In terms of indoor localization, the measurement model relates the obtained measurements from different sensors to the user location. In terms of proximity, the measurement model relates the obtained RSSI values from different sensors to the estimated RSSI values. 
\end{itemize}
Below, we formulate the indoor localization problem as a non-linear Bayesian tracking problem as used in our prior work and \cite{faheemsensys} and \cite{faheemglobecom}. 
\subsubsection{Indoor localization as Non-Linear Bayesian Tracking}
We model the indoor localization problem as a non-linear Bayesian tracking problem. We use RSSI and rate of change of RSSI as the state in \cite{faheemsensys}, while in this paper, our state consists of: a) Cartesian coordinates in 2D and 3D environments for PF-based tracking; and  b) RSSI and the rate of change of RSSI for RSSI smoothing.    
For indoor localization, let the state (in our case state is RSSI value for KF and user position for a PF) sequence $x_i$ $\{x_i, i \;\epsilon \; \mathbb{N} \}$ of a given target be represented by Equation \ref{eq:sysmodel} in the system model.
\begin{equation}
x_i=f_i(x_{i-1},v_{i-1}) 
\label{eq:sysmodel}
\end{equation}
The function $f_i$ is a non-linear state function (which can be linear as well depending on the system or state dynamics) that relates the current position (for PF) or RSSI value (for KF) $x_i$ to the previous position (PF) or RSSI value (KF) $x_{i-1}$ and the process noise sequence $v_{i-1}$. The motive behind tracking is to recursively calculate the user's location using measurements (obtained from iBeacons in our case) as given in Equation \ref{eq:measmod}. $h_i$ is a non-linear function (can be linear as well based on requirement) that relates the obtained measurement with the estimated user position  (or RSSI for proximity detection). $w_i$ is the measurement noise sequence. 

\begin{equation}
z_i=h_i(x_{i},w_{i}) 
\label{eq:measmod}
\end{equation}
The Bayesian tracking (or proximity detection problem) is fundamentally  recursively calculating some belief in the estimated user location (or RSSI values) $x_i$ at time $i$ based on the obtained measurements $z_{1:i}$ up to time \textit{i}. Therefore, it is important to obtain the probability density function (pdf) $p(x_i|z_{1:i})$. 
We assume that the initial pdf $p(x_0|z_0)$ 
 is the same as the prior probability $p(x_0)$ of the position vector. We recursively calculate $p(x_{i}|z_{1:i-1})$ using the \textit{prediction}, and \textit{update} stages discussed below:
\begin{itemize}
	\item \textit{Prediction}: 
	We assume that we have the required pdf $p(x_{i-1}|z_{1:i-1})$ at time $i-1$. We use the system model in Equation \ref{eq:sysmodel} to calculate the prior position pdf at time $i$ using Chapman-Kolmogorov equation as done by Arulampalam et al \cite{arulampalam2002tutorial}. 
	\begin{equation}
	p(x_i|z_{1:i-1}) = \int p(x_i|x_{i-1}) p(x_{i-1}|z_{1:i-1})dx_{i-1}
	\label{eq:chap}
	\end{equation}
	\item \textit{Update}: Since measurements from sensors (iBeacons in our case) become available at the $i^{th}$ step, we update the prior in the update stage using Bayes' rule 
	given in Equation \ref{eq:uda}.
	\begin{equation}
	p(x_i|z_{1:i}) =  \frac{p(z_i|x_i)p(x_i|z_{1:i-1})}{p(z_i|z_{i-1})} 
	\label{eq:uda}
	\end{equation}
	where the normalizing constant $p(z_i|z_{i-1})$ is given by Equation \ref{eq:nor} and depends on $p(z_i|x_i)$, as defined in the measurement model given in Equation \ref{eq:measmod}. In the update stage, the measurements obtained from sensors are used to update the prior probability density to obtain the posterior probability density of user's current location. 
	\begin{equation}
	p(z_i|z_{i-1})= \int {p(z_i|x_i)p(x_i|z_{i-1})dx_i}
	\label{eq:nor}
	\end{equation}
\end{itemize}
The prediction and update stages allow us to obtain the optimal Bayesian solution \cite{arulampalam2002tutorial}. However, the time complexity of obtaining an optimal solution is not feasible in terms of the necessary computational time  for our indoor localization and proximity detection problem. Therefore, we use a PF to approximate the optimal solution for indoor localization and a KF for RSSI smoothing. Below, we present a short discussion on KF and PF. 
\subsubsection{Kalman Filter}
A KF is based on the assumption that the posterior probability distribution at every time instance $i$ is Gaussian that has a mean value $(\mu)$, and variance $(\sigma^2)$. Furthermore, if the process and measurement noise are Gaussian and both the state and obtained measurements evolve linearly with time,  then we obtain the following mathematical model, as described by Guvenc et al. \cite{guveenhancements}.

\begin{equation}
\label{eq:6}
x_i=Fx_{i-1}+v_i
\end{equation}
\begin{equation}
\label{eq:7}
z_i=H_i+w_i
\end{equation}
where $v_i \sim N(0,Q)$ and $w_i \sim N(0,R)$. $x_i$ represents the state vector, which in our case consists of RSSI and rate of change of RSSI (see Section 4 for details). $z_i$ represents the obtained measurements from the sensors (RSSI in our case). $Q$ is the process noise covariance and $R$ is the measurement noise covariance. Table \ref{tab:1} provides a list of the variables used in mathematical modelling of KF.   
\begin{table}
	\centering
	\caption{Kalman filter parameter notation}
	\label{tab:1}
	\begin{tabular}{|l| p{6cm}|}
		\hline
		\textbf{Symbol} & \textbf{Meaning}                 \\ \hline
		x      & State vector                     \\ \hline
		z      & Measurement/observation vector   \\ \hline
		F      & State transition matrix          \\ \hline
		P      & State vector estimate covariance or Error covariance \\ \hline
		Q      & Process noise covariance         \\ \hline
		R      & Measurement noise covariance     \\ \hline
		H     & Observation matrix               \\ \hline
		K & Kalman Gain \\\hline
		v & Process noise  \\ \hline
		w & Measurement noise \\ \hline
	\end{tabular}
\vspace{-12pt}
\end{table}

The prediction and update stages for the KF are:
\begin{enumerate}
	\item Prediction Stage: 
	\begin{equation}
	\label{eq:8}
	\hat{x}_{\bar{i}}= F\hat{x}_i
	\end{equation}
	\begin{equation}
	\label{eq:9}
	P_{\bar{i}}= FP_{i-1}F^T+Q
	\end{equation}
	where $P$ is the error covariance. 
	\item Update State:
	\begin{eqnarray}
	K_i=& P_{\bar{i}}H^T(HP_{\bar{i}}H^T+R)^{-1}\\
	\hat{x}_i=&\hat{x}_{\bar{i}}+K_i(z_i-H\hat{x}_{\bar{i}})\\
	P_i= &(I-K_iH)P_{\bar{i}}
	\end{eqnarray}
	The Kalman gain $K$ controls the reliance of the model on the obtained measurements. The higher the Kalman gain, the greater would be the reliance on the obtained measurements. 
\end{enumerate}

\begin{figure}[h!]
	\centering
	\includegraphics[width=0.46\textwidth]{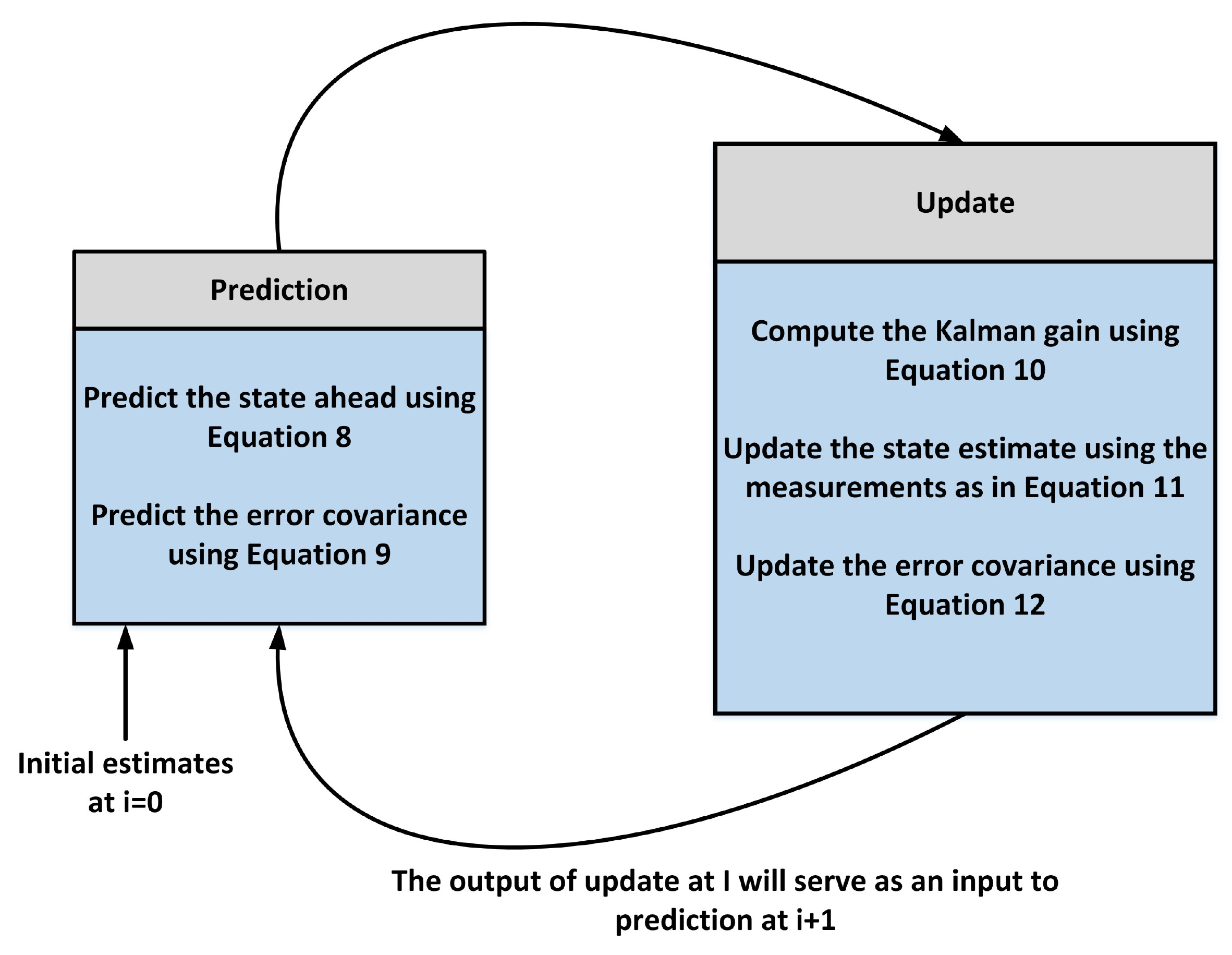}
	\caption{Prediction and update steps in Kalman Filter}
	\protect\label{fig:6}
	\vspace{-12pt}
\end{figure}

 The prediction and update are part of a recursive process as shown in the Figure \ref{fig:6}.
\subsubsection{Particle Filter}
A PF is a Monte Carlo (MC) method that is widely used for localization \cite{gustafsson2002particle,arulampalam2002tutorial}. Using a  PF, the posterior pdf is represented using random weighted samples that are used to compute an estimate of the state (user location in our case). With an increase in number of samples, the PF approaches the optimal estimate of the state. 
\par Mathematically, we characterize the posterior probability density function $p(x_{0:i}|z_{1:i})$ with set of random measures $ \left\{x_{0:i}^k, w_{i}^k\right\}$. The set $\left\{ x_{0:i}^k, k=0,.....N_s\right\}$ represents the support points set that has weights given by $\left\{ w_{i}^k, k=0...N_s\right\}$, while the state up to time $i$ is $x_{0:i}\left\{ x_{j}, j=0,.....i\right\}$. The weights are normalized and the posterior pdf at time $i$ is approximated by 
\begin{equation}
p(x_{0:i}|z_{1:i}) \approx \sum\limits_{k=1}^{N_s} w_i^k\delta(x_{0:i}-x_{0:i}^k)
\end{equation}
which is a discretely weighted approximation of the true posterior pdf.  The weights are chosen using Importance sampling (see \cite{bergman1999recursive}) 
\begin{equation}
w_i^k \propto \frac{p(x_{0:i}^k|z_{1:i})}{q(x_{0:i}^k|z_{1:i})}
\end{equation}
whereas $q(x_{0:i}|z_{1:i})$ represents the importance density. 
\par Due to the sequential nature of the problem, the obtained samples approximate  $p(x_{0:i-1}|z_{1:i-1})$ at every iteration. Therefore, a new sample set is needed to approximate $p(x_{0:i}^k|z_{1:i})$.  The importance density is given by 
\begin{equation}
q(x_{0:i}^k|z_{1:i}) = q(x_i|x_{0:i-1},z_{1:i})q(x_{0:i-1}|z_{1:i-1})
\end{equation}
The existing samples $x_{0:i}^k \sim q(x_{0:i-1}|z_{1:i-1})$ are augmented with the new state $x_i^k \sim q(x_i|x_{0:i-1},z_{1:i})$ to obtain the new samples $x_{0:i}^k \sim q(x_{0:i}|x_{1:i})$. The weight update equation is (see Arulampalam et al. \cite{arulampalam2002tutorial} for details)
\begin{equation}
x_i^k = w_{i-1}^k \frac{p(z_i|x_i^k)p(x_i^k|x_{i-1}^k) }{q(x_i^k|x_{0:i-1}^k,z_{1:i})}
\end{equation}
Similarly, $p(x_i|z_{1:i})$ is approximated as 
\begin{equation}
p(x_i|z_{1:i}) \approx \sum\limits_{k=1}^{N_s} w_{i}^k \delta(x_i-x_i^k)
\end{equation}
Hence, the PF algorithm is based on recursive propagation of particles and weights when the measurements are obtained sequentially.
\section{iBeacon-based Proximity and Indoor Localization System}
\label{sec:proxandlocalsys}
\subsection{Improving the proximity detection accuracy}
\subsubsection{Server-Side Running Average}
 In our first algorithm, \textit{Server-side Running Average}, we collect the RSSI values from the beacons using the user device and report them to a server. Rather than using RSSI directly as a measure of the user's proximity to any specific beacon, we relate it with distance using the path-loss model as described by Kumar in \cite{kumar2009distance} and given in Equation \ref{eq:pathloss}. In this equation, \textit{n} represents path-loss exponent that varies in value depending on the environment, \textit{d} is the distance between the user and the beacon, $d_0$ is the reference distance which is 1 meter in our case, while \textit{C} is the average RSSI value at $d_0$.
 \begin{equation}
 \label{eq:pathloss}
 RSSI=-10\; n\; log_{10}(d/d_0)+C
 \end{equation}
 Once the path-loss model is obtained, it efficiently characterizes the behavior of RSSI at different distances resulting in an accurate distance estimate. We believe that using a path-loss model that reflects the characteristics of the environment will improve the proximity detection accuracy as compared with the current approach. To account for the drastic fluctuations in the RSSI, the user is classified in a proximity zone only if three consecutive measurements obtained from the iBeacons classify him in that position through the estimated distance obtained using path-loss model. Algorithm 1 shows the SRA algorithm. 
\begin{algorithm}[!h]
	\small
	\caption{Server-side Running Average}
	\label{alg1}
	\begin{algorithmic}[1]
		\Procedure{Server-side Running Average}{}
		\State Obtain a path-loss model $P_L$ using site survey
		\State $D_0 \leftarrow 0$ \Comment{Initial distance}
		\State $P_0 \leftarrow Unknown$ \Comment{Initial proximity}
		\State Load $RSSI_{recv}$ \Comment{RSSI values received from sensors}
		\State $RSSI_{filt} \leftarrow RSSI_{recv} $ \Comment{iOS filtered RSSI values}
		\State $D_i \leftarrow D_0$ \Comment{Distance at sample \textit{i}}
		\State $P_i \leftarrow P_0$ \Comment{Proximity at sample \textit{i}}
		\State $P \leftarrow P_0$ \Comment{Classified proximity}
		\While{$RSSI_{filt} \neq 0$}
		\State  $D_i \leftarrow P_L(RSSI_{filt})$
		\State $P_i \leftarrow Proximity(D_i)$ \Comment{Zones using Table \ref{tab:proximityzones}}
		\If{$P_i$ is in zone \textit{x} for i= \textit{t-2,t-1,t}}
		\State $P \leftarrow P_i$
		\Else{ $P \leftarrow P$}
		\EndIf
		\EndWhile
		\EndProcedure
	\end{algorithmic}
\end{algorithm}
\subsubsection{Server-side Kalman Filter}
\begin{figure*}[t]
	\centering
	\includegraphics[width=0.8\textwidth]{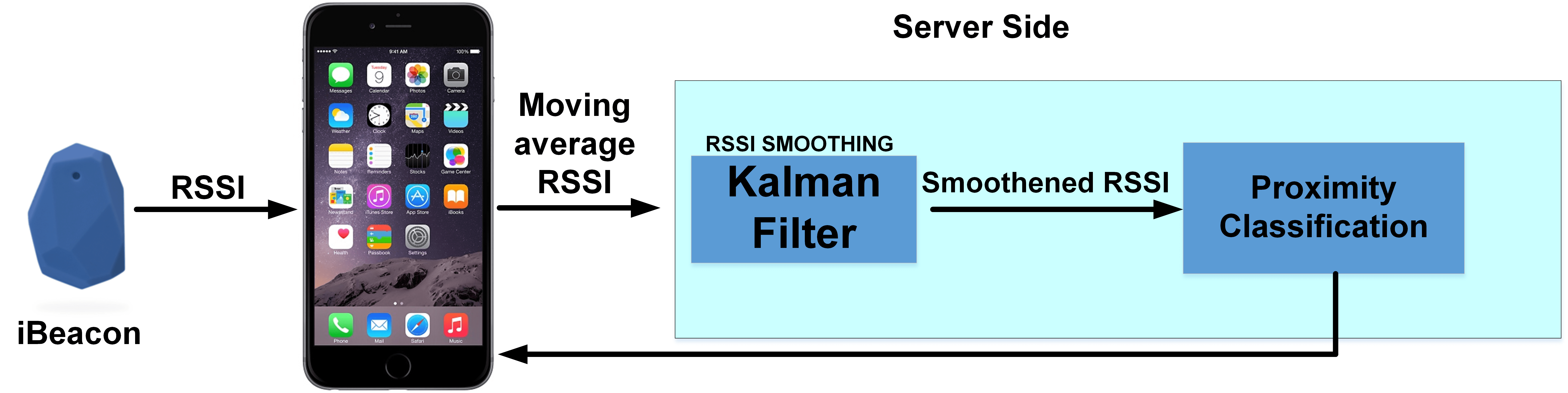}
	\caption{Proposed Kalman filter-based proximity detection}
	\protect\label{fig:5}
\vspace{-12pt}
\end{figure*}

Our second algorithm, \textit{Server-side Kalman Filter}, is  a modified version of SRA and utilizes Kalman Filtering to reduce the fluctuation in the RSSI as shown in Figure \ref{fig:5}. Since proximity is a mere estimation of location rather than exact position, we chose a Kalman filter over a particle filter due to reduced complexity. Using this approach, RSSI values from the beacons are received by the user device which are then forwarded to the  server that uses Kalman filtering to reduce signal  fluctuations. The smoothed RSSI values are then converted into a distance value using the path-loss model. Like SRA, the proximity is reported in a particular zone only if three consecutive measurements obtained from the iBeacons indicate the proximity of the user to the beacon to be in that zone. This means that the proximity decision for any specific samples depends on the two samples preceding it.  Our Kalman filter based RSSI smoother is based on the work of Guvenc \cite{guveenhancements}. 

For the purpose of filtering the RSSI values, we utilize a state vector $x_i$ that consists of the RSSI value $y_i$ and the rate of change of RSSI $\Delta y_{i-1}$ as given below. 
\begin{equation*}
x_i=\begin{bmatrix}
y_i\\ 
\Delta y_i
\end{bmatrix}
\end{equation*}
$\Delta y_i$ is dependent on the environment and signifies how drastically RSSI value fluctuates. The higher the noise in the environment, the higher will be the fluctuation. The current value of RSSI $y_i$  is assumed to be the previous RSSI $y_{i-1}$  plus the change $\Delta y_i$ and process noise $v_i^y$. Hence Equation \ref{eq:6} can be written as 
\begin{equation}
\label{eq:13}
\begin{bmatrix}
y_i\\ 
\Delta y_i
\end{bmatrix}
= \begin{bmatrix}
1 & \delta t \\ 
0 & 1 
\end{bmatrix}
\begin{bmatrix}
y_{i-1}\\ 
\Delta y_{i-1}
\end{bmatrix}
+
\begin{bmatrix}
v_i^{y}\\ 
v_i^{\Delta y}
\end{bmatrix} 
\end{equation}
which means that the state transition matrix F is given by 
\begin{equation*}
F=\begin{bmatrix}
1 &\delta t \\ 
0 & 1 
\end{bmatrix}
\end{equation*}
The parameter $\delta t$ is to be adjusted as per the variation in RSSI which depends on the environment. For our set of experiments,  $\delta t$ was taken as 0.2 (using trial and error). Similarly, Equation \ref{eq:7} can be rewritten as 
\begin{equation}
\label{eq:14}
\begin{bmatrix}
z_i
\end{bmatrix}
= \begin{bmatrix}
1 & 0  

\end{bmatrix}
\begin{bmatrix}
y_{i}\\ 
\Delta y_{i}
\end{bmatrix}
+
\begin{bmatrix}
w_i^{y}
\end{bmatrix} 
\end{equation}
The observation matrix H is given by 
\begin{equation*}
H=\begin{bmatrix}
1 & 0  
\end{bmatrix}
\end{equation*}
Parameters P, Q and R used in the experiments were obtained using trial and error, and are given below.
\begin{equation*}
\label{eq:15}
P =\begin{bmatrix}
100 & 0\\ 
0 & 100
\end{bmatrix}
Q = \begin{bmatrix}
0.001 & 0\\ 
0 & 0.001
\end{bmatrix}
R =
\begin{bmatrix}
0.10 
\end{bmatrix} 
\end{equation*}
The Kalman filter, once calibrated, effectively smooths the RSSI values. The smoothed RSSI values were then input into the path-loss model to obtain distances between the iBeacons and the user, and the user's proximity to the beacon was classified in any of the aforementioned zones. Algorithm 2 shows SKF. 
\begin{algorithm}[]
	\small
	\caption{Server-side Kalman Filter}
	\label{alg2}
	\begin{algorithmic}[1]
		\Procedure{Server-side Kalman Filter}{}
		\State Obtain a path-loss model $P_L$ using site survey
		\State $D_0 \leftarrow 0$ \Comment{Initial distance}
		\State $P_0 \leftarrow Unknown$ \Comment{Initial proximity}
		\State Load $RSSI_{recv}$ \Comment{RSSI values received from sensors}
		\State $RSSI_{filt} \leftarrow RSSI_{recv} $ \Comment{iOS filtered RSSI values}
		\State $D_i \leftarrow D_0$ \Comment{Distance at sample \textit{i}}
		\State $P_i \leftarrow P_0$ \Comment{Proximity at sample \textit{i}}
		\State $P \leftarrow P_0$ \Comment{Classified proximity}
		\While{$RSSI_{filt} \neq 0$}
		\State $RSSI_{filt} \leftarrow KalmanFilter(RSSI_{filt}) $ 
		\State  $D_i \leftarrow P_L(RSSI_{filt})$
		\State $P_i \leftarrow Proximity(D_i)$ \Comment{Zones using Table \ref{tab:proximityzones}}
		\If{$P_i$ is zone \textit{x} for i= \textit{t-2,t-1,t}}
		\State $P \leftarrow P_i$
		\Else{ $P \leftarrow P$}
		\EndIf
		\EndWhile
		\EndProcedure
	\end{algorithmic}
\end{algorithm}

\subsection{iBeacon-based Indoor Localization}
\begin{figure}[h!]
	\centering
	\includegraphics[width=0.46\textwidth]{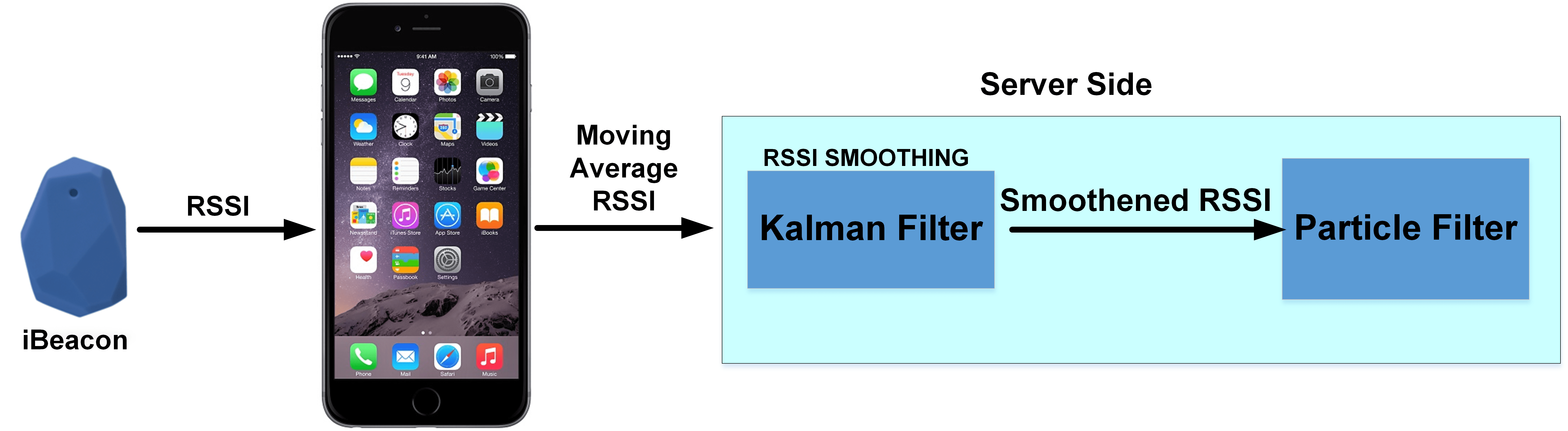}
	\caption{KFPF approach used by the server-side iBeacon-based indoor localization system}
	\protect\label{fig:kfpffig}
	\vspace{-12pt}
\end{figure}

We implement a KFPF cascade approach for improving the localization accuracy of RSSI-based indoor localization system compared with the PF and KF approximations singularly used to an optimal Bayesian solution. We achieve the improvement in accuracy using a 2 step approach.
\begin{enumerate}
	\item \textit{Use KF to smooth the RSSI values}
	\item \textit{Use smoothed RSSI values as input to PF for non-linear tracking}
\end{enumerate}
The filtered RSSI values, are converted into distance using a path-loss model given in Equation \ref{eq:pathloss}.
The distance estimates, obtained from RSSI values, are then used by the PF algorithm to estimate the user location as discussed in Section \ref{3} using the Non-Linear Bayesian tracking model discussed in Section 3.1. Algorithm \ref{alg3} describes our proposed approach. Figure \ref{fig:kfpffig} shows the KFPF approach used by server-side iBeacon-based indoor localization system. 

\begin{algorithm}
	\caption{Kalman Filter-Particle Filter (KFPF) Cascade}
	\label{alg3}
	\begin{algorithmic}[1]
		\Procedure{KFPF Cascade}{}
		\State Load $RSSI_{recv}$ \Comment{RSSI values received from sensors}
		\State $RSSI \leftarrow RSSI_{recv}$
		\State $RSSI_{filt} \leftarrow 0$ \Comment{Filtered RSSI values}
		\State $L_i \leftarrow (0,0) or (0,0,0)$ \Comment{Estimated 2D or 3D user location}
		\While{$RSSI \neq 0$}
		\State $RSSI_{filt} \leftarrow KalmanFilter(RSSI)$
		\State  $L_i \leftarrow ParticleFilter(RSSI_{filt})$
		\State Print $L_i$
		\EndWhile
		\State \textbf{\textit{end}}
		\EndProcedure
	\end{algorithmic}
\end{algorithm}
\vspace{-12pt}

\section{Experimental Results and Discussions}
\label{sec:expres}
\subsection{iBeacon based Indoor Proximity Services}
\begin{figure}
	\centering
	\includegraphics[width=0.46\textwidth]{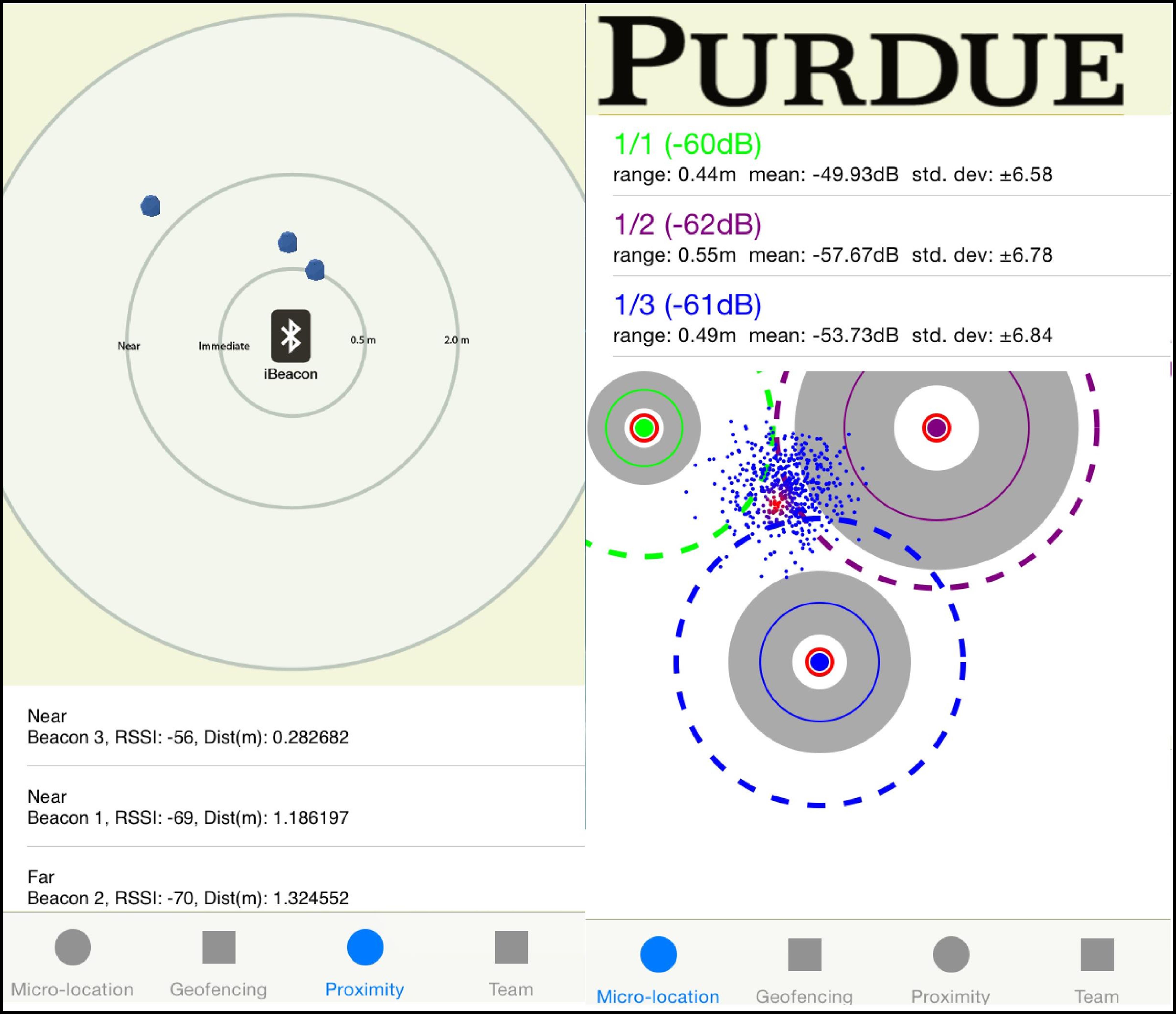}
	\caption{Our prototype iOS application. }
	\protect\label{fig:3a}
	\vspace{-12pt}
\end{figure}

To evaluate the proximity detection accuracy of our two algorithms, we placed a Gimbal \cite{gimbal} beacon in two different rooms (for cross validation) which are 11m $\times$ 6m (environment 1) and 8m $\times$ 4m (environment 2) in dimension. The rooms, due to the infrastructure inside, replicate a typical real world scenario in which beacons are utilized.  We used an iPhone 6s plus running the latest iOS version 9.2 as the user device. We conducted experiments with iPhone 4s as well, however there was no significant difference so we proceeded with iPhone 6s plus. The iPhone was loaded with our prototype application that can be seen in Figure \ref{fig:3a}\footnote{Figure \ref{fig:3a} is taken from our prior work \cite{faheemglobecom}}. Our prototype application primarily has two main functionalities; micro-location (used in our prior work \cite{faheemglobecom}) and proximity-based services. The left side of the Figure \ref{fig:3a}  shows the proximity tab. This application can listen to several beacons in its vicinity and then classify them into locations based on the RSSI value.  The right hand side of the Figure \ref{fig:3a}  shows the micro-location tab.
A core-i5 Macbook-pro with 8 gigabytes of RAM, running  Apache Tomcat 8.0 and Java 1.8 was used as the server to run the SRA and SKF algorithms. Table \ref{tab:sum} summarizes the equipment related information used in the tests. 

\begin{table}[h!]
	\centering
	\caption{Summary of Device Parameters}
	\begin{tabular}{|l|l|}
		\hline
		Server & Apache Tomcat\\\hline
		Java version & Java 1.8 \\\hline
		User Device  & Apple iPhone 6s plus  \\ \hline
		Wireless Interface & Bluetooth V4.2 / 2.4GHz \\ \hline
		Operating System & iOS 9.2          \\ \hline
		Beacons   & Gimbal Series 10 \\ \hline
		Gimbal range & 50 meters \\ \hline
		Transmission Frequency & 100 ms           \\ \hline
		Major Value  & Yes              \\ \hline
		Minor Value  & Yes              \\ \hline
	\end{tabular}
	\label{tab:sum}
\end{table}
\par
\begin{figure}[t!]
\centering
\includegraphics[width=0.5\textwidth]{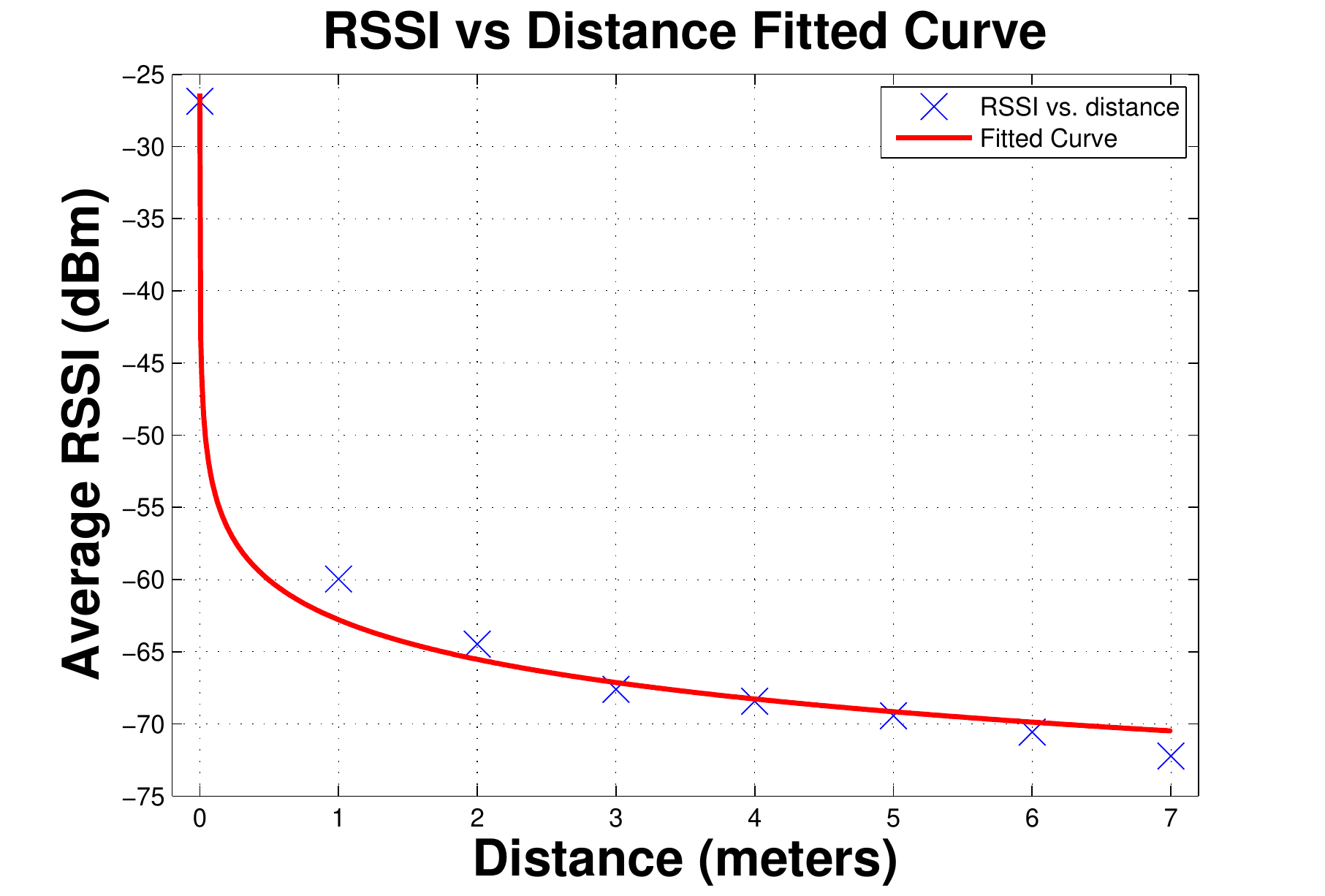}
\caption{Curve fitting for RSSI values at distances from 0 to 7 meters in Environment 1} \label{fig:6a}
\vspace{-12pt}
\end{figure}

\begin{figure}
\includegraphics[width=0.5\textwidth]{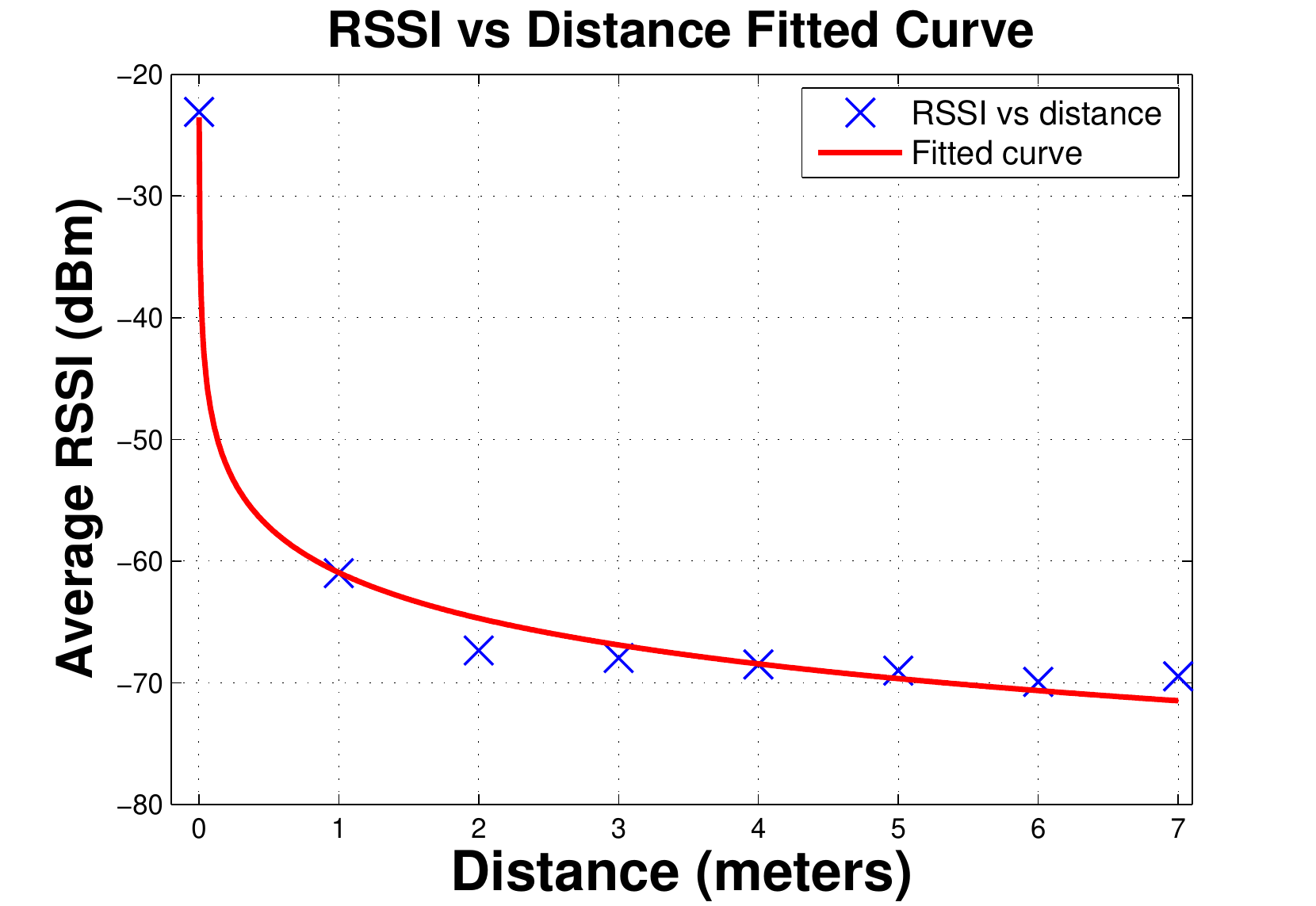}
\caption{Curve fitting for RSSI values at distances from 0 to 7 meters in Environment 2} \label{fig:6b}
\vspace{-12pt}
\end{figure}

 To obtain the path-loss models for our environments, we put the beacon in a fixed position and noted the average RSSI values on a user device for a number of distances starting from 0 meter up to 7 meters.  We collected and averaged 22 RSSI samples at each location. As shown in Figure \ref{fig:6a} and \ref{fig:6b} respectively, we plotted distance vs. average RSSI and then used Matlab's curve fitting function to estimate a curve for distance vs RSSI in both environment 1 and environment 2.
The fitted curve for environment 1, based on Equation \ref{eq:pathloss}, has path-loss exponent \textit{n} equal to  0.9116 with 95\% confidence bounds between (0.8272, 0.996) and \textit{C} (where C is the RSSI value at a reference distance $d_0$ taken to be 1 meter for iBeacons) equal to -62.78  with 95\% confidence bounds between (-64.07, -61.05) while the $R^2$ value for the fitted curve is 0.9915. For environment 2, the path-loss exponent \textit{n} equals to  1.246 with 95\% confidence bounds between (1.139, 1.354) and \textit{C} equals to -60.95  with 95\% confidence bounds between (-62.24, -59.66) while the $R^2$ value for the fitted curve is 0.9926.
Using the above values in Equation \ref{eq:pathloss}, we obtained Equation \ref{eq:pathloss2} for environment 1.  Equation \ref{eq:pathloss2} is rearranged into Equation \ref{eq:distance} to obtain the distance from the beacons using RSSI values in environment 1.  Similarly for environment 2,  Equation \ref{eq:pathloss3} is rearranged into Equation \ref{eq:distance2} to obtain the distance from the beacons using RSSI values.
Table \ref{tab:2} and Table \ref{tab:2b} list the average RSSI values at different distances from the beacons along with the actual distance, computed distance and the estimation error for environment 1 and 2 respectively. 
\begin{equation}
\label{eq:pathloss2}
RSSI=-10 \times 0.9116 \times log_{10}d-62.78
\end{equation}
\begin{equation}
\label{eq:distance}
d=10^{(\frac{62.78+RSSI}{-9.116})}
\end{equation}
\begin{equation}
\label{eq:pathloss3}
RSSI=-10 \times 1.246 \times log_{10}d-60.95
\end{equation}
\begin{equation}
\label{eq:distance2}
d=10^{(\frac{60.95+RSSI}{-12.46})}
\end{equation}

\begin{table}[]
\centering
\caption{An insight into the estimation error of the fitted curve for environment 1}
\label{tab:2}
\begin{tabular}{|l|p{1.8cm}|p{1.4cm}|p{1cm}|}
\hline
\textbf{Average RSSI} & \textbf{Actual Distance (m)} & \textbf{Computed Distance (m)} &\textbf{Error (m)} \\ \hline
-26.8692    & 0.0001           & 0.0001                              & 0                 \\ \hline
-59.9565	              & 1                            & 0.4901                              & 0.5099                  \\ \hline
-64.4782
& 2                          & 1.5357                             & 0.4643                  \\ \hline
-67.6086             & 3                           & 3.3861                             & 0.3861                  \\ \hline
-68.4347              & 4                           & 4.1717                             & 0.1717                  \\ \hline
-69.4347              & 5                            & 5.3705                             & 0.3705                 \\ \hline
-70.5652              & 6                           & 7.1452                              & 1.1452                  \\ \hline
-72.2173
& 7                           & 10.8457                             & 3.8457                  \\ \hline
\end{tabular}
\end{table}
\begin{table}[]
\centering
\caption{An insight into the estimation error of the fitted curve for environment 2}
\label{tab:2b}
\begin{tabular}{|l|p{1.8cm}|p{1.4cm}|p{1cm}|}
\hline
\textbf{Average RSSI} & \textbf{Actual Distance (m)} & \textbf{Computed Distance (m)} & \textbf{Error (m)} \\ \hline
-23.1034          & 0.0001           & 0.0009              & 0.0008         \\ \hline
-61                   & 1               & 1.0093                & 0.0093         \\ \hline
-67.3448          & 2               & 3.2601                & 1.2601         \\ \hline
-67.9655          & 3               & 3.6563                & 0.6563         \\ \hline
-68.5         & 4               & 4.0359                & 0.0359         \\ \hline
-69                  & 5               & 4.4266                & 0.5734         \\ \hline
-69.9310          & 6               & 5.2576                & 0.7424         \\ \hline
-69.4827         & 7               & 4.8396                & 2.1604         \\ \hline
\end{tabular}
\end{table}

\par Using these models, we evaluated the performance of SRA and SKF and used the current approach of moving averaging of RSSI values as the benchmark.  To evaluate the performance of the current approach, we put the beacon in a fixed position and noted the estimated proximity at different distances. During our experiments, we tested the models only in the `immediate', `near' and `far' regions since the `unknown' region is of no practical use. We obtained the user's proximity using the three different approaches at a distance of 0, 0.6, 1.8, 2.4, 4.3, and 5.5 meters. These distances, in contrast with Table \ref{tab:2} and \ref{tab:2b}, are chosen such that we have two evaluation points in each proximity zone. 
We took 20 RSSI samples at each physical location where each RSSI sample consists of the running average of 10 RSSI values (Apple reports RSSI after every 1 second while our gimbal beacon transmitted every 100ms). For every proximity zone, we took 40 samples (20 samples $\times$ 2 distances)  resulting in 120 (3 zones $\times$ 40 samples for each zone) RSSI samples for each approach.
\par
\begin{figure}[t!]
\centering
\includegraphics[width=0.5\textwidth]{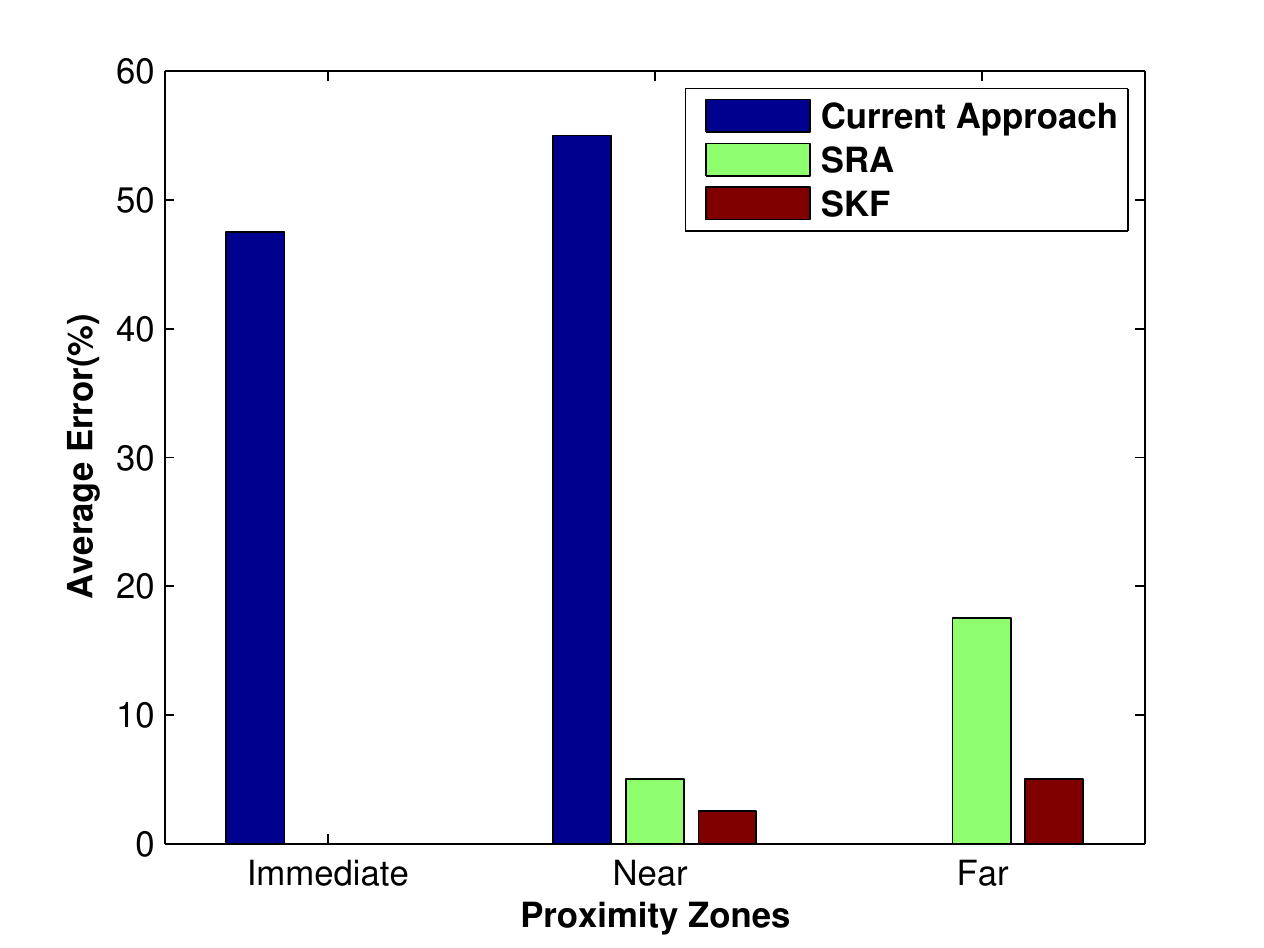}
\caption{Error in different proximity zones for the models in Environment 1} \label{fig:7a}
\vspace{-15pt}
\end{figure}

\begin{figure}
\includegraphics[width=0.5\textwidth]{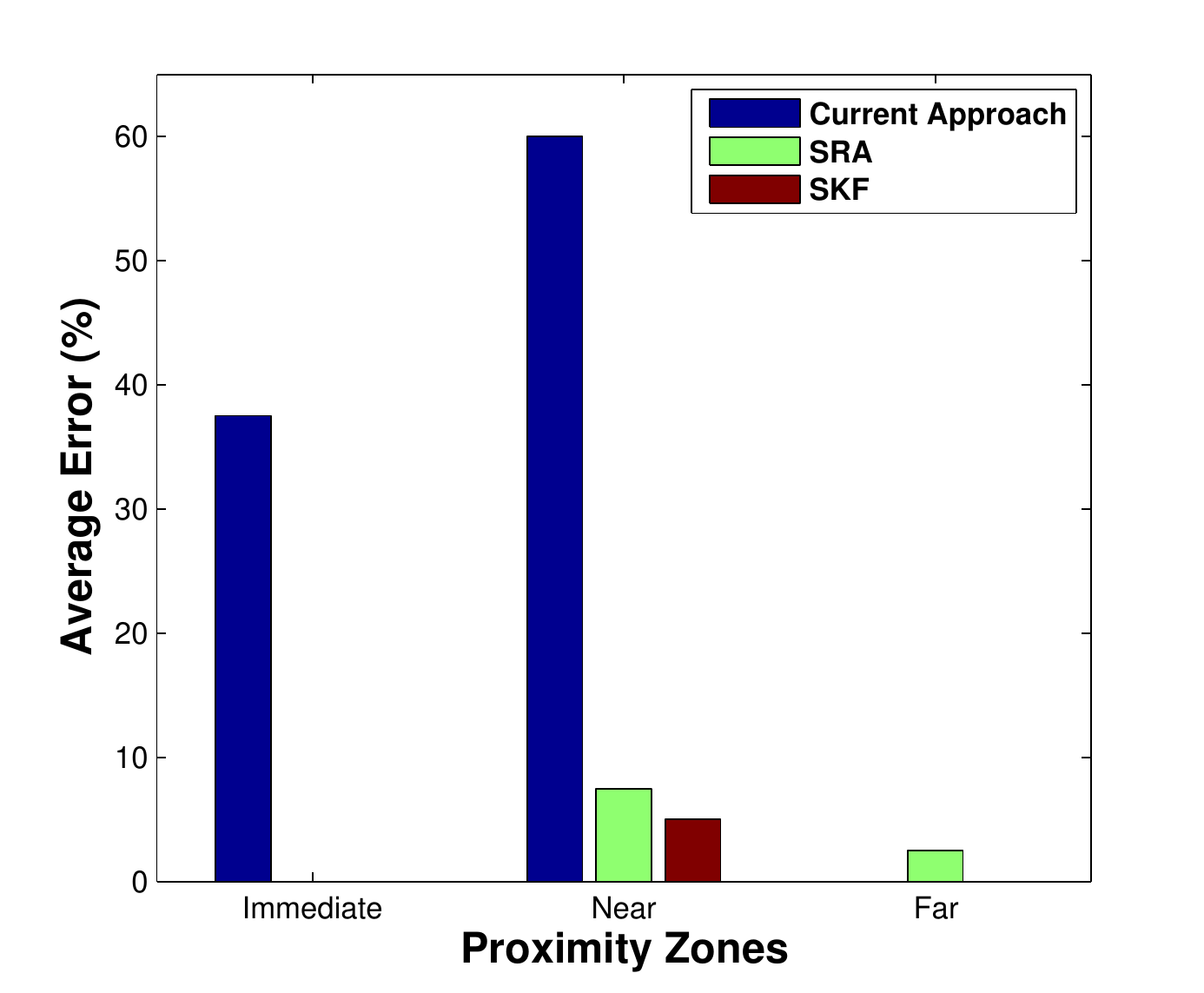}
\caption{Error in different proximity zones for the models in Environment 2} \label{fig:7b}
\vspace{-15pt}
\end{figure}

 To compare SRA, SKF, and the current approach, we used a three class \textit{confusion matrix} which is one of the popular methods used for evaluating the performance of classification models \cite{patro2015novel}. The matrix compares the actual classification with the predicted or estimated classification. Table \ref{tab:confusionmatrix} lists the parameters of the utilized confusion matrix as described by Fawcett in \cite{fawcett2006introduction} and provides their description in context of our experiment.  Table \ref{tab:conres} and \ref{tab:conres2} contain confusion matrix statistics obtained for SRA and SKF along with the current approach used as the benchmark for both environment 1 and environment 2 respectively.  
Table \ref{tab:est} and \ref{tab:est2} show the proximity error at different distances in different proximity zones for environment 1 and environment 2 respectively. The tables also show the total proximity error in every proximity zone. Figure \ref{fig:7a} and \ref{fig:7b} show the error for the three approaches in the `immediate', `near' and `far' zones for both environments. The SRA and SKF have 0\% error in the `immediate' zone compared to the 47.5\%  and 37.5\% observed using the current approach in environment 1 and environment 2 respectively. 
 
\begin{figure}[!h]
\centering
\includegraphics[width=0.5\textwidth]{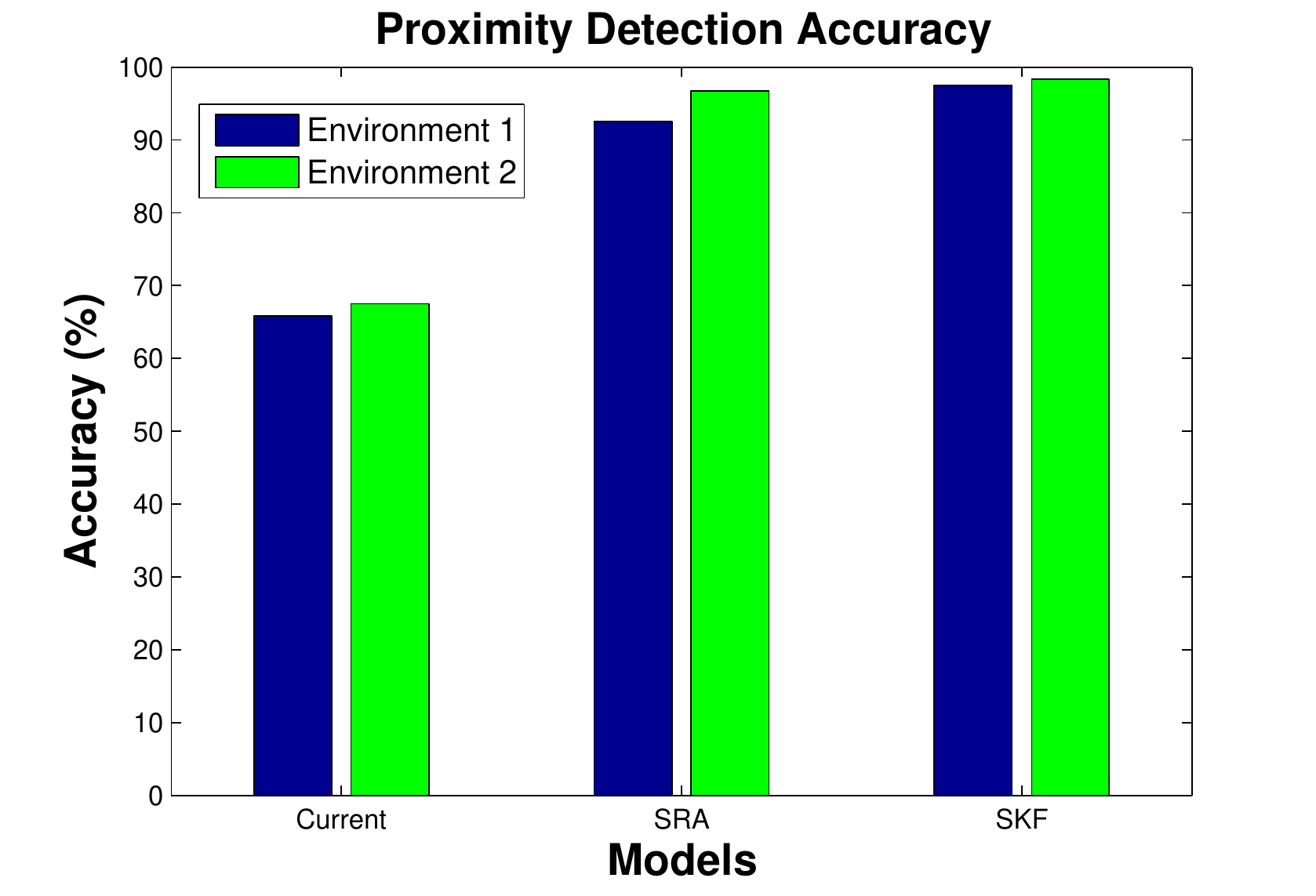}
\caption{Average Proximity Detection Accuracy of three algorithms}
\protect\label{fig:9}
\end{figure}
\begin{table*}[t!]
\centering
\caption{Different parameters used in confusion matrix}
\label{tab:confusionmatrix}
\begin{tabular}{|p{3.0cm}|p{13cm}|}
\hline
\textbf{Parameter} & \textbf{Description} \\ \hline
True Positive (TP)	&  When the user is in zone `\textit{x}' and is classified in zone `\textit{x}' \\ \hline
True Negative (TN)	 & When the user is not in zone `\textit{x}' and is not classified in zone `\textit{x}' \\ \hline
False Positive (FP)	& When the user is not in zone `\textit{x}' but is classified in zone `\textit{x}'\\ \hline
False Negative (FN)	& When the user is in zone `\textit{x}' but is not classified in zone `\textit{x}' \\ \hline
Precision/ Positive Prediction Value (PPV)	&  The fractions of samples classified in zone `\textit{x}'. Mathematically, precision = $\frac{TP_i}{TP_i+FP_i}$ where $i$ is any zone.\\ \hline
Sensitivity/Recall	&  The fraction of samples correctly classified in zone `\textit{x}'. Mathematically,  sensitivity = $\frac{TP_i}{TP_i+FN_i}$. The higher the sensitivity, the better will be the algorithm. \\ \hline
Specificity	& The fraction of samples correctly classified in any zone other than zone `\textit{x}'. The higher the specificity, the better will be the algorithm. Mathematically, specificity = $\frac{TN_i}{TN_i+FP_i}$.  \\ \hline
Fall out/False Positive Rate (FPR) & Mathematically, FPR = 1-specificity = $\frac{FP_i}{FP_i+TN_i}$. The lower the FPR value, the better will be the algorithm.   \\ \hline
False Negative Rate (FNR) & Mathematically, FNR = $\frac{FN_i}{FN_i+TP_i}$. The lower the FNR value, the better will be the algorithm. \\ \hline
False Discovery Rate (FDR) & A good indicator for conceptualizing the rate of type I error. Mathematically, FDR = 1 - PPV = $\frac{FP_i}{FP_i+TP_i}$ . The lower the FDR value, the better will be the algorithm.  \\ \hline
Accuracy & The fraction of samples correctly classified. Mathematically, accuracy = $\frac{TP+TN}{TP+TN+FP+FN} \forall$  zones.  \\ \hline
\end{tabular}
\end{table*} 
\begin{table*}[htb!]
\centering
\caption{Statistical metrics for the current approach, SRA and SKF in environment 1}
\label{tab:conres}
\begin{tabular}{|l|l|l|l|l|l|l|l|l|l|}
\hline
\multicolumn{1}{|c|}{\multirow{2}{*}{\textbf{Metrics}}} & \multicolumn{3}{l|}{\textbf{Immediate}} & \multicolumn{3}{l|}{\textbf{Near}} & \multicolumn{3}{l|}{\textbf{Far}} \\ \cline{2-10} 
\multicolumn{1}{|c|}{}                                  & \textbf{Current} & \textbf{SRA} & \textbf{SKF} & \textbf{Current} & \textbf{SRA} & \textbf{SKF} & \textbf{Current} & \textbf{SRA} & \textbf{SKF} \\ \hline
\textbf{True Positive}                                             & 21                   & 40           & 40           & 18                   & 38           & 39           & 40                   & 33           & 38           \\ \hline
\textbf{True Negative}                                             & 80                   & 78           & 79           & 61                   & 73           & 78           & 58                   & 80           & 80           \\ \hline
\textbf{False Positive}                                             & 0                    & 2            & 1            & 19                   & 7            & 2            & 22                   & 0            & 0            \\ \hline
\textbf{False Negative}                                             & 19                   & 0            & 0            & 22                   & 2            & 1            & 0                    & 7            & 2            \\ \hline
\textbf{Precision}                                      & 1                    & 0.952  & 0.975  & 0.486          & 0.844     & 0.951  & 0.645           & 1            & 1            \\ \hline
\textbf{Sensitivity}                                    & 0.525                & 1            & 1            & 0.45                 & 0.95         & 0.975        & 1                    & 0.825        & 0.95         \\ \hline
\textbf{Specificity}                                    & 1                    & 0.975        & 0.987       & 0.762               & 0.912       & 0.975        & 0.725                & 1            & 1            \\ \hline
\textbf{Fall out}                                       & 0                    & 0.025        & 0.012       & 0.237               & 0.087       & 0.025        & 0.275                & 0            & 0            \\ \hline
\textbf{FDR}                                            & 0                    & 0.047  & 0.024  & 0.513        & 0.155     & 0.048  & 0.354           & 0            & 0            \\ \hline
\textbf{False Negative Rate}                            & 0.475                & 0            & 0            & 0.55                 & 0.05         & 0.025        & 0                    & 0.175        & 0.05         \\ \hline
\end{tabular}
\end{table*}

\begin{table*}[tbh!]
\centering
\caption{Statistical metrics for the current approach, SRA and SKF in environment 2}
\begin{tabular}{|l|l|l|l|l|l|l|l|l|l|}
\hline
\multicolumn{1}{|c|}{\multirow{2}{*}{\textbf{Metrics}}} & \multicolumn{3}{l|}{\textbf{Immediate}}            & \multicolumn{3}{l|}{\textbf{Near}}                 & \multicolumn{3}{l|}{\textbf{Far}}                  \\ \cline{2-10} 
\multicolumn{1}{|c|}{}                                  & \textbf{Current} & \textbf{SRA} & \textbf{SKF} & \textbf{Current} & \textbf{SRA} & \textbf{SKF} & \textbf{Current} & \textbf{SRA} & \textbf{SKF} \\ \hline
\textbf{True Positive}   & 25 & 40  & 40  & 16   & 37  & 38   & 40  & 39     & 40   \\ \hline
\textbf{True Negative}   & 80   & 80 & 80  & 65   & 79   & 80  & 56  & 77  & 78           \\ \hline
\textbf{False Positive}    & 0  & 0   & 0  & 15    & 1  & 0 & 24 & 3  & 2 \\ \hline
\textbf{False Negative}  & 15  & 0   & 0   & 24   & 3   & 2      & 0   & 1 & 0 \\ \hline
\textbf{Precision}    & 1  & 1  & 1  & 0.516   & 0.973 & 1  & 0.625  & 0.928  & 0.952            \\ \hline
\textbf{Sensitivity}                                    & 0.625                & 1            & 1            & 0.4                 & 0.925         & 0.95        & 1                    & 0.975        & 1        \\ \hline
\textbf{Specificity}   & 1                    & 1        & 1       & 0.812               & 0.987       & 1        & 0.7                & 0.962            & 0.975            \\ \hline
\textbf{Fall out}                                       & 0                    & 0        & 0      & 0.187               & 0.012       & 0        & 0.3           & 0.037            & 0.025            \\ \hline
\textbf{FDR}                                            & 0                    & 0  & 0  & 0.483       & 0.026    & 0  & 0.375           & 0.071            & 0.047            \\ \hline
\textbf{False Negative Rate}     & 0.375 & 0  & 0  & 0.6 & 0.075   & 0.05   & 0                    & 0.025        & 0        \\ \hline
\end{tabular}
\label{tab:conres2}
\end{table*}
\begin{table*}[hbt!]
\centering
\caption{Comparison of proximity detection error of SRA and SKF in comparison with current approach in environment 1}
\label{tab:est}
\begin{tabular}{|l|l|l|l|l|l|l|l|}
\hline
\multirow{2}{*}{\textbf{Actual}}    & \multirow{2}{*}{\textbf{Distance (meters)}} & \multicolumn{3}{l|}{\textbf{\centering Error at Different Distances (\%)}}                         & \multicolumn{3}{l|}{\textbf{Error in Different Zones (\%)}}   \\ \cline{3-8} 
&                                             & \textbf{Current} & \textbf{SRA} & \textbf{SKF} & \textbf{Current}   & \textbf{SRA}    & \textbf{SKF}    \\ \hline
\multirow{2}{*}{\textbf{Immediate}} & 0                                           & 0                    & 0                & 0                 & \multirow{3}{*}{47.5} & \multirow{3}{*}{0}     & \multirow{3}{*}{0}     \\ \cline{2-5}
& 0.6     & 95   & 0    & 0    &     &   &       \\ \hline
\multirow{2}{*}{\textbf{Near}}      & 1.8                                         & 10                   & 10               & 5               & \multirow{3}{*}{55} & \multirow{3}{*}{5} & \multirow{3}{*}{2.5}  \\ \cline{2-5}
& 2.4   & 100  & 0                 & 0                &                        &     &                                \\ \hline
\multirow{3}{*}{\textbf{Far}}       & 4.3                                         & 0                 & 5          & 0               & \multirow{2}{*}{0}  & \multirow{3}{*}{17.5}    & \multirow{3}{*}{5} \\ \cline{2-5}& 5.5                                           & 0                  & 30                & 10                 &                        &                        &                           \\ \hline
\end{tabular}
\end{table*}
\begin{table*}[htb!]
\centering
\caption{Comparison of proximity detection error of SRA and SKF in comparison with current approach in environment 2}
\label{tab:est2}
\begin{tabular}{|l|l|l|l|l|l|l|l|}
\hline
\multirow{2}{*}{\textbf{Actual}}    & \multirow{2}{*}{\textbf{Distance (meters)}} & \multicolumn{3}{l|}{\textbf{\centering Error at Different Distances (\%)}}                         & \multicolumn{3}{l|}{\textbf{Error in Different Zones (\%)}}                         \\ \cline{3-8} 
&  & \textbf{Current} & \textbf{SRA} & \textbf{SKF} & \textbf{Current}   & \textbf{SRA}    & \textbf{SKF}    \\ \hline
\multirow{2}{*}{\textbf{Immediate}} & 0                                           & 0                    & 0                & 0                 & \multirow{3}{*}{37.5} & \multirow{3}{*}{0}     & \multirow{3}{*}{0}     \\ \cline{2-5}
& 0.6                                         & 75                  & 0                & 0                 &                        &                        &                      \\ \hline
\multirow{2}{*}{\textbf{Near}}      & 1.8                                         & 20                  & 0              & 0             & \multirow{3}{*}{60} & \multirow{3}{*}{7.5} & \multirow{3}{*}{5.0}  \\ \cline{2-5}
& 2.4                                           & 100                & 15                 & 10                &                        &                        &                                \\ \hline
\multirow{3}{*}{\textbf{Far}}       & 4.3                                         & 0                 & 5          & 0               & \multirow{2}{*}{0}  & \multirow{3}{*}{2.5}    & \multirow{3}{*}{0} \\ \cline{2-5}
& 5.5                                           & 0                  & 0                & 0                 &                        &                        &                           \\ \hline
\end{tabular}
\end{table*}
Figure \ref{fig:9} shows the average proximity detection accuracy of the two proposed approaches in both environments in comparison with the current approach used as the benchmark. Our algorithms, SRA and SKF, outperform Apple's current approach for proximity detection both in environment 1 and environment 2.
\par As mentioned earlier, the first step in our approach was to obtain the path-loss models that accurately reflected the noise characteristics of our utilized environments. Figures \ref{fig:6a} and \ref{fig:6b} show that our curve-fitted path-loss model in environment 1 and 2 respectively can accurately estimate the distance between any user and beacon using the RSSI values.  The $R^2$ value of 0.9915 and 0.9926 highlight the accuracy of the fitted models. These results are also confirmed by Table \ref{tab:2} and \ref{tab:2b}. The average error between the actual distance and estimated distance is 86.14 cm and 67.98 cm for environment 1 and environment 2 respectively.  The results in Table \ref{tab:conres} and \ref{tab:conres2} highlight the improvements that are attained using SRA and SKF in comparison with the current approach.  The higher value of the true positives and true negatives for both SRA and SKF indicates that our algorithms can accurately detect the user's location in any particular zone. The lower value of false positives and false negatives for both SRA and SKF in comparison with the current approach means that our algorithms do not falsely detect or reject a user in a particular zone. Similarly, the sensitivity values for SRA and SKF are higher than the current approach in both the `immediate' and `near' zones. The higher sensitivity values mean that both SRA and SKF are more sensitive and able to detect the user in a particular zone as compared to the current approach. SKF performs better than SRA as well.  The improved proximity detection of SRA and SKF is also highlighted by the lower FDR, FNR, and FPR values. The high FNR value for the current approach in the `immediate' and `near' zone means that the current approach is not suitable for these zones. Furthermore, the high FDR value for the current approach in the `far' zone means that it is more likely to incorrectly classify the user's proximity in the `far' zone. This is why the zero error in `far' zone for current approach, given in Table \ref{tab:est} and \ref{tab:est2}, is not significant as it is due to the inherent flaw in the current approach. \par 
In Table \ref{tab:est} and \ref{tab:est2}, a proximity error of 95\% and 75\% respectively for the current approach at 0.6 meters (which falls in the immediate region) means that for 19 (environment 1) and 15 (environment 2)  out of 20 samples collected at this distance, the current approach was not able to accurately classify them into the `immediate' zone. The average error of 47.5\% and 37.5\% for the current approach in the `immediate' region of environment 1 and 2 respectively, means that for the 40 samples collected in this region, only 21 samples in environment 1 and 25 samples in environment 2 were correctly classified. This shows the current approach is not favorable for PBS. SRA and SKF, on the other hand, have 100\% accuracy in the `immediate' zone in both environments. In the `near' region, error for the current approach is 55\% and 60\% for environment 1 and 2 respectively while it is 5\% and 2.5\% for SRA and SKF respectively in environment 1 and 7.5\% and 5\% in environment 2. This means that out of 40 samples, the current approach accurately classified only 18 samples and 16 samples in environment 1 and environment 2 respectively, which is far less than the 38 and 39 samples accurately classified by SRA and SKF respectively in environment 1 and 37 and 38 samples accurately classified by SRA and SKF respectively in environment 2.\par  Figure \ref{fig:7a} and \ref{fig:7b} show the average error for all three approaches. It can be seen that our SRA and SKF approaches outperform the current approach in the `immediate' and `near' zones, which is primarily used for triggering PBS in most of the beacon-based applications. The current approach, due to the inherent flaw of not taking the environmental factors into account, classifies most of the samples in the `far' region, which is why there is no detection error in the `far' zone for the current approach (the high FDR value in Table \ref{tab:conres} and \ref{tab:conres2} proves this fact). This is also the cause of higher detection error in the `immediate' and `near' regions for the current approach. As shown in Figure \ref{fig:9}, the current approach achieved a proximity detection accuracy of 65.83\% and 67.5\% in environment 1 and environment 2 respectively.  SRA achieved 92.5\% and 96.6\% proximity detection accuracy which is  26.7\% and 29.1\% improvement over the current approach in environment 1 and 2 respectively. SKF achieved a proximity detection accuracy of 97.5\% and 98.3\%, which improves proximity detection accuracy by almost 31.6\% and 30.8\% over the current approach in environment 1 and environment 2 respectively. The figure indicates that out of 120 samples, only 79 and 81 were properly classified by the current approach in environment 1 and environment 2 respectively, which is much lower than the 111 and 116 correctly classified by our approach SRA in environment 1 and environment 2 respectively, and 117 and 118 samples correctly classified by our approach SKF in environment 1 and environment 2 respectively. The increased accuracy of both SRA and SKF in two different environments makes it a viable alternative to the current approach. The improved performance of current approach, SRA and SKF in environment 2, when compared with environment 1, is due to less noise in environment 2 as evident from Figure \ref{fig:6b} and lower values of $C$ in Equation \ref{eq:pathloss3}.  

\subsection{iBeacon based Indoor Localization}
\begin{figure}[h!]
\centering
\includegraphics[width=0.46\textwidth]{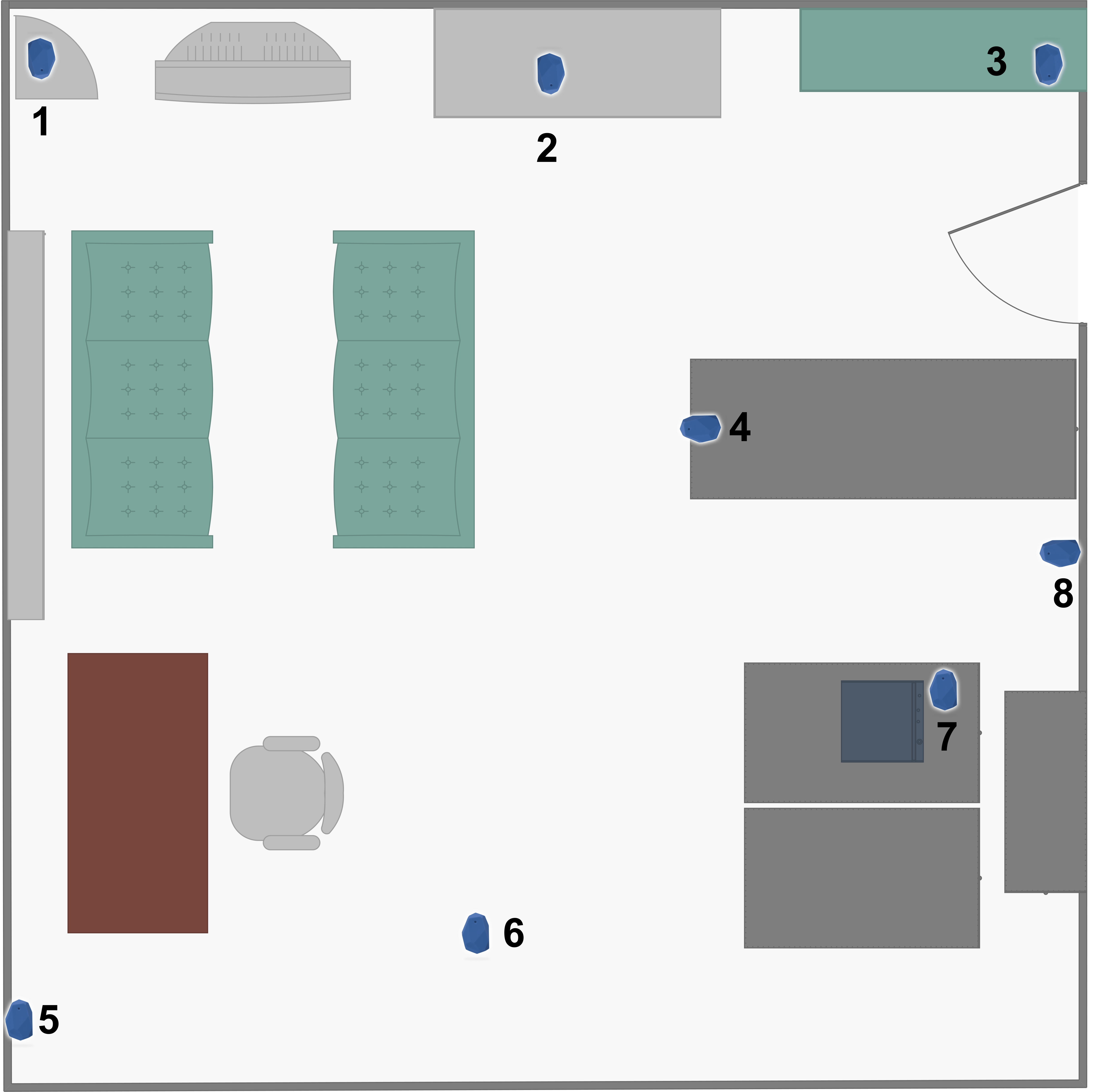}
\caption{The floor plan of the experiment space with 8 iBeacons. }
\protect\label{fig:3}
\vspace{-12pt}
\end{figure}

To evaluate the performance of our proposed cascaded filter approach, we chose an area of 7m $\times$ 6m and deployed a variable number of iBeacons for the experiments as shown in Figure \ref{fig:3}. The experimental space contained different obstacles that replicated the real world scenario. 
Table \ref{tab:sum} lists the device parameters used in our experiment. The server ran the PF and KFPF algorithms to obtain user location based on the RSSI values. 
We implemented the algorithm on the server following the thin-client paradigm, i.e. we moved the intensive computations to the server to optimize battery consumption of the user device. However, the cascaded filter approach can also run on user devices.  
We evaluated the  2D localization error using Equation \ref{eq:error} and 3D localization error using Equation \ref{eq:error1}.

\begin{table*}[ht!]
\centering
\caption{2D Localization error $E_{2D}$ (meters) for a varying number of particles and iBeacons using Particle Filter}
\begin{tabular}{|l|l|l|l|l|l|l|l|l|l|l|l|l|}
\hline
\multicolumn{1}{|c|}{\multirow{2}{*}{\textbf{Particles}}} & \multicolumn{2}{c|}{\textbf{3 Beacons}} & \multicolumn{2}{c|}{\textbf{4 Beacons}} & \multicolumn{2}{c|}{\textbf{5 Beacons}} & \multicolumn{2}{c|}{\textbf{6 Beacons}} & \multicolumn{2}{c|}{\textbf{7 Beacons}} & \multicolumn{2}{c|}{\textbf{8 Beacons}} \\ \cline{2-13} 
\multicolumn{1}{|c|}{}                                    & \textbf{Mean}   & \textbf{Std}  & \textbf{Mean}   & \textbf{Std}  & \textbf{Mean}   & \textbf{Std}  & \textbf{Mean}   & \textbf{Std}  & \textbf{Mean}   & \textbf{Std}  & \textbf{Mean}   & \textbf{Std}  \\ \hline
\textbf{400}                                              & 1.776           & 0.743         & 1.566           & 0.586         & 1.672           & 1.027         & 1.405           & 0.628         & 0.916           & 0.431         & 1.247           & 0.483         \\ \hline
\textbf{600}                                              & 1.768           & 0.641         & 1.788           & 0.881         & 1.693           & 1.084         & 1.307           & 0.589         & 1.063           & 0.533         & 1.395           & 0.641         \\ \hline
\textbf{800}                                              & 1.733           & 0.565         & 1.639           & 0.669         & 1.545           & 0.821         & 1.314           & 0.456         & 0.954           & 0.386         & 1.545           & 0.762         \\ \hline
\textbf{1000}                                             & 1.696           & 0.668         & 1.658           & 0.721         & 1.551           & 0.773         & 1.243           & 0.486         & 0.955           & 0.428         & 1.438           & 0.701         \\ \hline
\textbf{1200}                                             & 1.724           & 0.719         & 1.688           & 0.748         & 1.548           & 0.742         & 1.267           & 0.493         & \textbf{0.859}  & 0.382         & 1.506           & 0.664         \\ \hline
\textbf{1400}                                             & 1.665           & 0.724         & 1.704           & 0.561         & 1.502           & 0.624         & 1.252           & 0.470         & 0.995           & 0.391         & 1.346           & 0.403         \\ \hline
\textbf{1600}                                             & 1.701           & 0.570         & 1.703           & 0.879         & 1.547           & 0.984         & 1.259           & 0.530         & 0.959           & 0.454         & 1.448           & 0.654         \\ \hline
\textbf{1800}                                             & 1.643           & 0.498         & 1.647           & 0.605         & 1.932           & 0.674         & 1.300           & 0.639         & 1.017           & 0.392         & 1.304           & 0.647         \\ \hline
\textbf{2000}                                             & 1.681           & 0.557         & 1.715           & 0.655         & 1.526           & 0.925         & 1.278           & 0.512         & 1.010           & 0.339         & 1.225           & 0.585         \\ \hline
\end{tabular}
\label{tab:pf}
\end{table*}

\begin{table*}[th!]
\centering
\caption{2D Localization error $E_{2D}$ (meters) for a varying number of particles and iBeacons using Kalman Filter-Particle Filter}
\begin{tabular}{|l|l|l|l|l|l|l|l|l|l|l|l|l|}
\hline
\multicolumn{1}{|c|}{\multirow{2}{*}{\textbf{Particles}}} & \multicolumn{2}{c|}{\textbf{3 Beacons}} & \multicolumn{2}{c|}{\textbf{4 Beacons}} & \multicolumn{2}{c|}{\textbf{5 Beacons}} & \multicolumn{2}{c|}{\textbf{6 Beacons}} & \multicolumn{2}{c|}{\textbf{7 Beacons}} & \multicolumn{2}{c|}{\textbf{8 Beacons}} \\ \cline{2-13} 
\multicolumn{1}{|c|}{}                                    & \textbf{Mean}   & \textbf{Std}  & \textbf{Mean}   & \textbf{Std}  & \textbf{Mean}   & \textbf{Std}  & \textbf{Mean}   & \textbf{Std}  & \textbf{Mean}   & \textbf{Std}  & \textbf{Mean}   & \textbf{Std}  \\ \hline
\textbf{400}                                              & 1.307           & 0.680         & 1.091           & 0.555         & 1.147           & 0.585         & 0.836           & 0.394         & 0.916           & 0.431         & 0.812           & 0.556         \\ \hline
\textbf{600}                                              & 1.484           & 0.745         & 1.219           & 0.425         & 1.013           & 0.532         & 0.902           & 0.369         & 0.985           & 0.502         & 0.849           & 0.705         \\ \hline
\textbf{800}                                              & 1.442           & 0.756         & 1.238           & 0.424         & 1.055           & 0.488         & 0.778           & 0.384         & 0.736           & 0.470         & 0.782           & 0.537         \\ \hline
\textbf{1000}                                             & 1.472           & 0.826         & 1.276           & 0.500         & 0.939           & 0.512         & 0.837           & 0.331         & 0.748           & 0.501         & 0.796           & 0.598         \\ \hline
\textbf{1200}                                             & 1.412           & 0.827         & 1.450           & 0.513         & 1.090           & 0.524         & 0.806           & 0.341         & 0.709           & 0.528         & 0.804           & 0.650         \\ \hline
\textbf{1400}                                             & 1.390           & 0.702         & 1.371           & 0.463         & 1.064           & 0.591         & 0.845           & 0.387         & 0.724           & 0.513         & 0.781           & 0.547         \\ \hline
\textbf{1600}                                             & 1.449           & 0.773         & 1.376           & 0.481         & 1.023           & 0.524         & 0.861           & 0.410         & 0.742           & 0.465         & 0.849           & 0.598         \\ \hline
\textbf{1800}                                             & 1.500           & 0.833         & 1.295           & 0.520         & 1.149           & 0.494         & 0.821           & 0.394         & 0.714           & 0.368         & 0.786           & 0.507         \\ \hline
\textbf{2000}                                             & 1.452           & 0.696         & 1.193           & 0.381         & 1.240           & 0.663         & 0.816           & 0.352         & \textbf{0.708}  & 0.382         & 0.819           & 0.496         \\ \hline
\end{tabular}
\label{tab:kfpf}
\end{table*}

\begin{figure}[h!]
\centering
\includegraphics[width=0.46\textwidth]{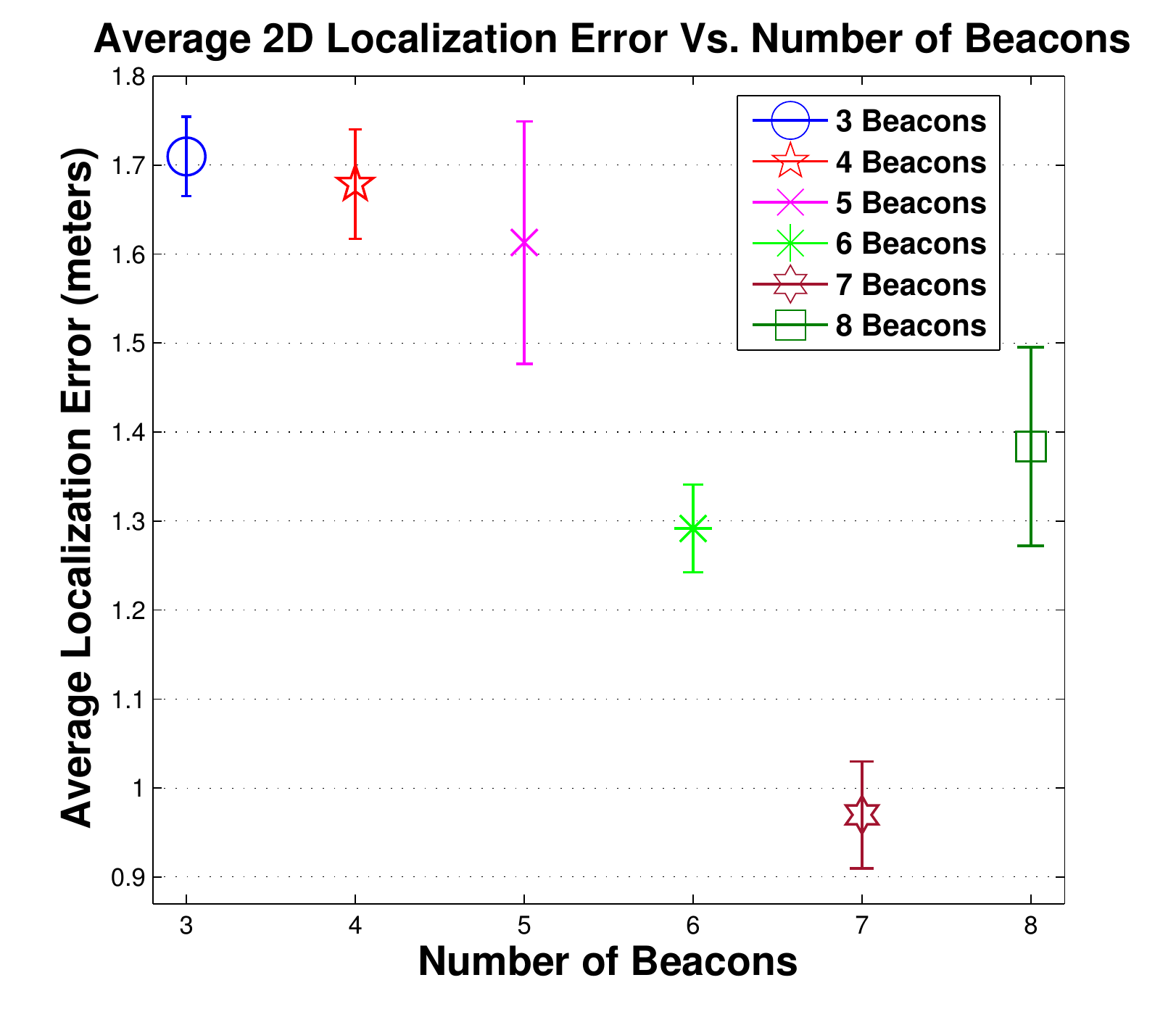}
\caption{Average 2D Localization Error $E_{2D}$ Vs Number of Particles for a Varying Number of Beacons for Particle filter algorithm.}
\protect\label{fig:pferrorbar}
\vspace{-12pt}
\end{figure}

\begin{figure}[h!]
\centering
\includegraphics[width=0.46\textwidth]{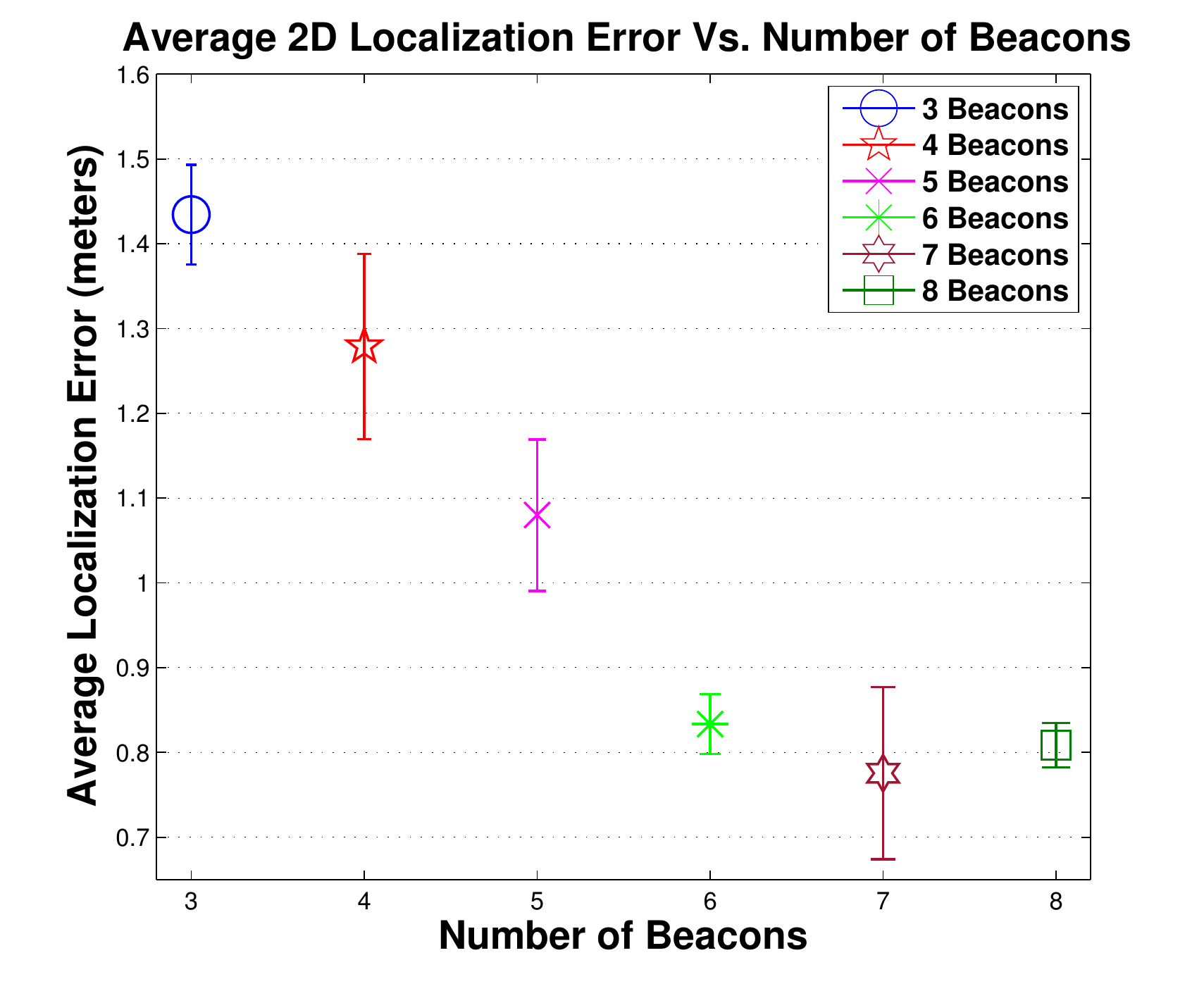}
\caption{Average 2D Localization Error $E_{2D}$ Vs Number of Particles for a Varying Number of Beacons for Kalman Filter-Particle filter algorithm. }
\protect\label{fig:kfpferrorbar}
\end{figure}

\begin{equation}
\label{eq:error}
E_{2D}= \frac{\sum_{i=1}^{n}\sqrt{(X_i-X_{<est>})^2+(Y_i-Y_{<est>})^2}}{n}
\end{equation} 
\begin{equation}
\label{eq:error1}
\begin{split}
E_{3D}= \frac{\sum_{i=1}^{n}\sqrt{(X_i-X_{<est>})^2+(Y_i-Y_{<est>})^2}}{n} \\
+ \frac{\sum_{i=1}^{n}\sqrt{(Z_i-Z_{<est>})^2}}{n}
\end{split}
\end{equation} 

\begin{figure}[h!]
\centering
\includegraphics[width=0.46\textwidth]{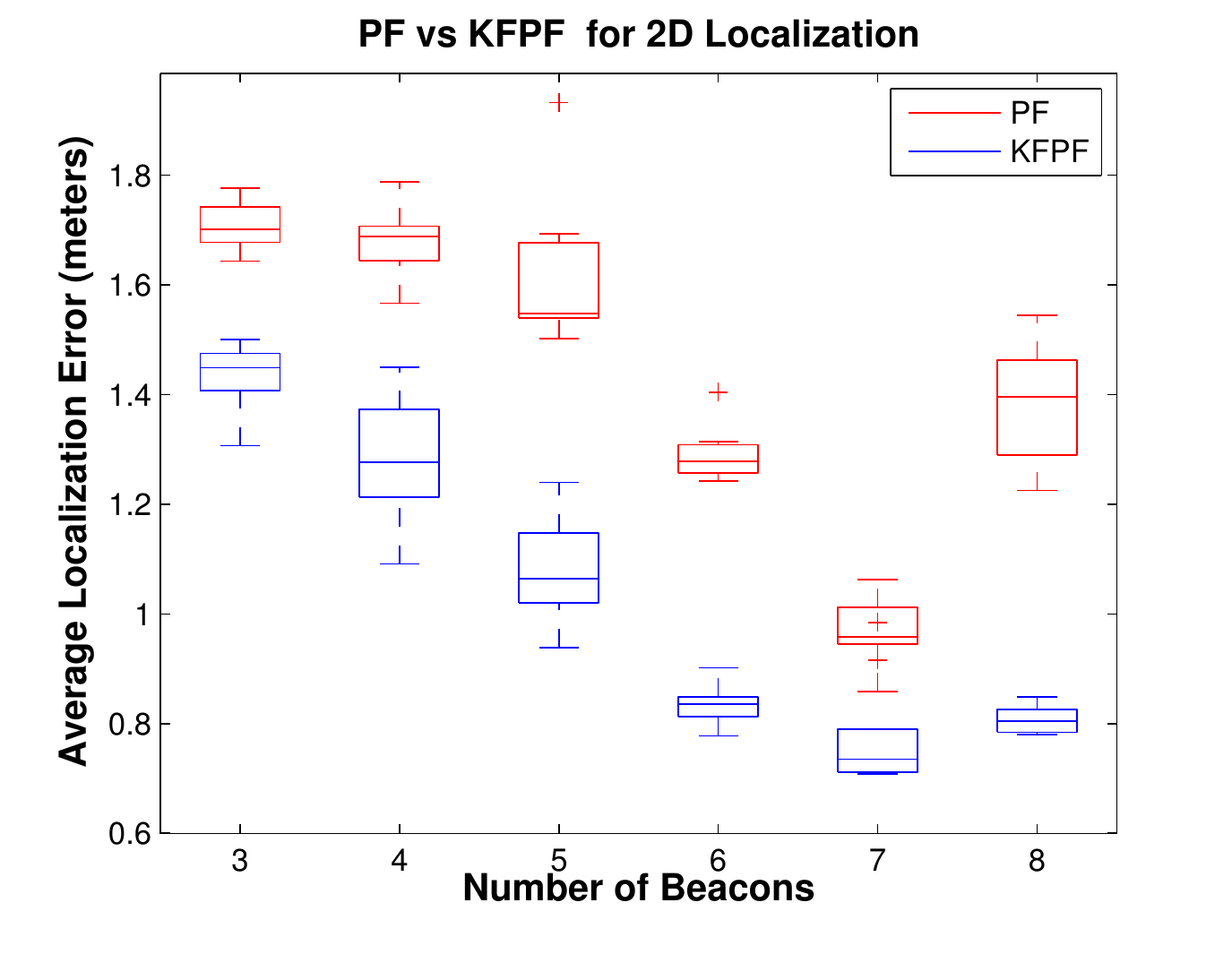}
\caption{Comparison of Particle Filter and Kalman Filter-Particle Filter for 2D localization. }
\protect\label{fig:boxplot}
\end{figure}
In Equation \ref{eq:error1}, $(X_i,Y_i,Z_i)$ is the actual point while the $(X_{<est>},Y_{<est>},Z_{<est>}) $ is the average estimated point. $n$ is the total number of samples collected for each measurement, and n=10 for our set of experiments. We altered the number of iBeacons and particles (the support points discussed in Section 3.3) in PF to evaluate the localization error. We increased the number of iBeacons from 3 to 8 during the experiment.  Below we discuss the localization results for 2D and 3D environment separately. 
\subsection{2D Localization Results}
Table \ref{tab:pf} shows the average 2D localization error $E_{2D}$ and standard deviation of $E_{2D}$ for a different number of iBeacons and particles using a Particle filter algorithm. The lowest 2D localization error of 0.859 meters was obtained using 7 beacons and 1200 particles. The average 2D localization error using the PF algorithm is 1.44 meters.  Figure \ref{fig:pferrorbar} shows that the 2D localization error was reduced with addition of iBeacons. The lowest 2D localization error was obtained using 7 iBeacons. However, the addition of an $8^{th}$ iBeacon increased the localization error compared to seven iBeacon system. This observation is in line with our prior work \cite{faheemglobecom}.  This is probably due to the self interference among the iBeacons caused by the saturation of the experiment space with iBeacons.
\par Table \ref{tab:kfpf} shows the average 2D localization error and the standard deviation of localization error for different number of iBeacons and particles using our KFPF cascaded filter based approach. The lowest localization error of 0.70 meters was obtained using seven iBeacons and 2000 particles. The average 2D localization error using our KFPF algorithm was 1.03 meters. This is a 28.16\% improvement over the PF algorithm.  Figure \ref{fig:kfpferrorbar} shows that the 2D localization error reduced with addition of iBeacons. The lowest localization error was obtained using seven iBeacons and 2000 particles. However, the addition of the $8^{th}$  iBeacon, like the PF, increased the localization error compared to the seven iBeacon system.   
\par Figure \ref{fig:boxplot} compares the PF and the KFPF approaches in terms of average 2D localization error for a different number of iBeacons. It can be seen that the KFPF outperforms the PF approach in all setups and the use of a Kalman filter for filtering the fluctuating RSSI values is a viable approach that can enhance the accuracy of the PF in general. The highest improvement is seen with six beacons.

\begin{table*}[t]
\centering
\caption{3D Localization error $E_{3D}$ (meters) for various number of particles and iBeacons using Particle Filter}
\begin{tabular}{|l|l|l|l|l|l|l|l|l|l|l|l|l|}
\hline
\multirow{2}{*}{\textbf{Particles}} & \multicolumn{2}{c|}{\textbf{3 Beacons}} & \multicolumn{2}{c|}{\textbf{4 Beacons}} & \multicolumn{2}{c|}{\textbf{5 Beacons}} & \multicolumn{2}{c|}{\textbf{6 Beacons}} & \multicolumn{2}{c|}{\textbf{7 Beacons}} & \multicolumn{2}{c|}{\textbf{8 Beacons}} \\ \cline{2-13} 
& \textbf{Mean}   & \textbf{Std}  & \textbf{Mean}   & \textbf{Std}  & \textbf{Mean}   & \textbf{Std}  & \textbf{Mean}   & \textbf{Std}  & \textbf{Mean}   & \textbf{Std}  & \textbf{Mean}   & \textbf{Std}  \\ \hline
\textbf{400}                        & 2.101           & 0.651         & 1.963           & 0.574         & 2.234           & 1.153         & 1.649           & 0.615         & 1.243           & 0.404         & 1.454           & 0.536         \\ \hline
\textbf{600}                        & 2.094           & 0.574         & 2.139           & 0.817         & 2.232           & 1.279         & 1.582           & 0.562         & 1.287           & 0.540         & 1.620           & 0.608         \\ \hline
\textbf{800}                        & 2.075           & 0.536         & 1.956           & 0.542         & 2.190           & 1.014         & 1.616           & 0.470         & 1.238           & 0.435         & 1.746           & 0.809         \\ \hline
\textbf{1000}                       & 2.016           & 0.619         & 2.009           & 0.617         & 2.228           & 1.064         & 1.572           & 0.448         & 1.266           & 0.448         & 1.623           & 0.672         \\ \hline
\textbf{1200}                       & 2.135           & 0.825         & 2.086           & 0.783         & 2.191           & 1.002         & 1.577           & 0.467         & \textbf{1.189}  & 0.418         & 1.689           & 0.646         \\ \hline
\textbf{1400}                       & 2.068           & 0.777         & 2.071           & 0.437         & 2.169           & 0.880         & 1.570           & 0.449         & 1.263           & 0.466         & 1.534           & 0.488         \\ \hline
\textbf{1600}                       & 2.176           & 0.820         & 2.062           & 0.776         & 2.169           & 1.187         & 1.537           & 0.529         & 1.203           & 0.474         & 1.611           & 0.700         \\ \hline
\textbf{1800}                       & 1.990           & 0.517         & 2.020           & 0.564         & 2.177           & 0.668         & 1.581           & 0.621         & 1.325           & 0.418         & 1.473           & 0.698         \\ \hline
\textbf{2000}                       & 2.000           & 0.553         & 2.084           & 0.585         & 2.139           & 1.097         & 1.570           & 0.545         & 1.321           & 0.449         & 1.429           & 0.649         \\ \hline
\end{tabular}
\label{tab:pf3d}
\end{table*}

\begin{table*}[t]
\centering
\caption{3D Localization error $E_{3D}$ (meters) for various number of particles and iBeacons using Kalman Filter-Particle Filter}
\begin{tabular}{|l|l|l|l|l|l|l|l|l|l|l|l|l|}
\hline
\multirow{2}{*}{\textbf{Particles}} & \multicolumn{2}{c|}{\textbf{3 Beacons}} & \multicolumn{2}{c|}{\textbf{4 Beacons}} & \multicolumn{2}{c|}{\textbf{5 Beacons}} & \multicolumn{2}{c|}{\textbf{6 Beacons}} & \multicolumn{2}{c|}{\textbf{7 Beacons}} & \multicolumn{2}{c|}{\textbf{8 Beacons}} \\ \cline{2-13} 
& \textbf{Mean}   & \textbf{Std}  & \textbf{Mean}   & \textbf{Std}  & \textbf{Mean}   & \textbf{Std}  & \textbf{Mean}   & \textbf{Std}  & \textbf{Mean}   & \textbf{Std}  & \textbf{Mean}   & \textbf{Std}  \\ \hline
\textbf{400}                        & 1.623           & 0.729         & 1.418           & 0.551         & 1.349           & 0.641         & 1.119           & 0.403         & 1.243           & 0.404         & \textbf{0.947}  & 0.533         \\ \hline
\textbf{600}                        & 1.821           & 0.765         & 1.571           & 0.472         & 1.284           & 0.634         & 1.152           & 0.420         & 1.100           & 0.513         & 1.084           & 0.686         \\ \hline
\textbf{800}                        & 1.731           & 0.792         & 1.594           & 0.408         & 1.314           & 0.617         & 1.041           & 0.454         & 1.065           & 0.518         & 1.009           & 0.566         \\ \hline
\textbf{1000}                       & 1.803           & 0.879         & 1.657           & 0.494         & 1.267           & 0.621         & 1.133           & 0.448         & 1.146           & 0.542         & 1.054           & 0.595         \\ \hline
\textbf{1200}                       & 1.704           & 0.812         & 1.823           & 0.467         & 1.391           & 0.620         & 1.079           & 0.432         & 1.074           & 0.550         & 1.043           & 0.667         \\ \hline
\textbf{1400}                       & 1.699           & 0.706         & 1.707           & 0.456         & 1.384           & 0.691         & 1.108           & 0.490         & 1.136           & 0.532         & 1.031           & 0.516         \\ \hline
\textbf{1600}                       & 1.694           & 0.787         & 1.687           & 0.466         & 1.308           & 0.643         & 1.111           & 0.513         & 1.099           & 0.545         & 1.051           & 0.632         \\ \hline
\textbf{1800}                       & 1.800           & 0.825         & 1.615           & 0.490         & 1.457           & 0.602         & 1.100           & 0.492         & 1.133           & 0.459         & 1.029           & 0.549         \\ \hline
\textbf{2000}                       & 1.740           & 0.700         & 1.619           & 0.374         & 1.505           & 0.734         & 1.141           & 0.523         & 0.948           & 0.480         & 1.090           & 0.561         \\ \hline
\end{tabular}
\label{tab:kfpf3d}
\end{table*} 

\begin{figure}[h!]
\centering
\includegraphics[width=0.46\textwidth]{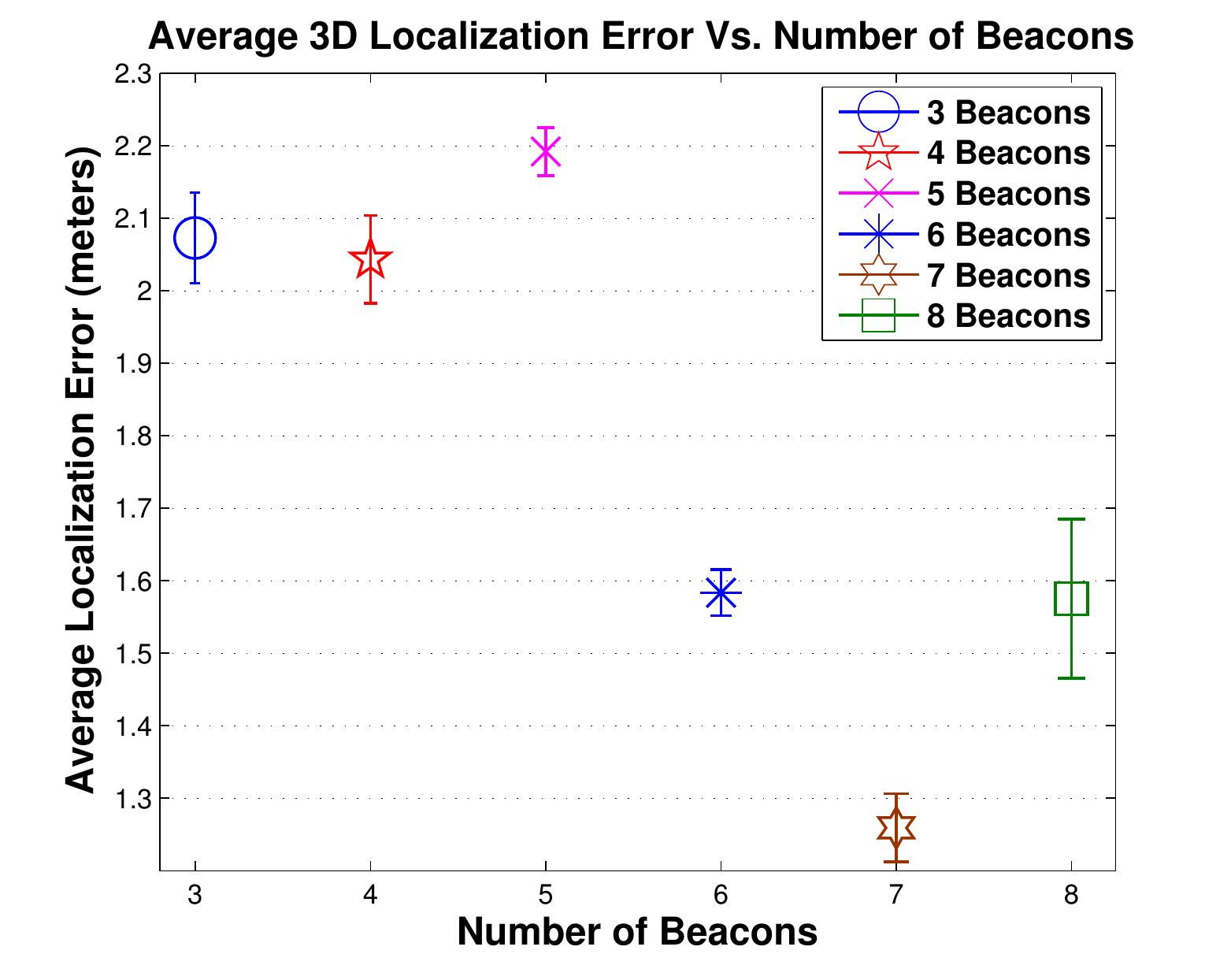}
\caption{Average 3D Localization Error $E_{3D}$ Vs Number of Particles for a varying Number of Beacons for Particle filter algorithm. }
\protect\label{fig:3dpf}
\end{figure}  
\begin{figure}[h!]
\centering
\includegraphics[width=0.46\textwidth]{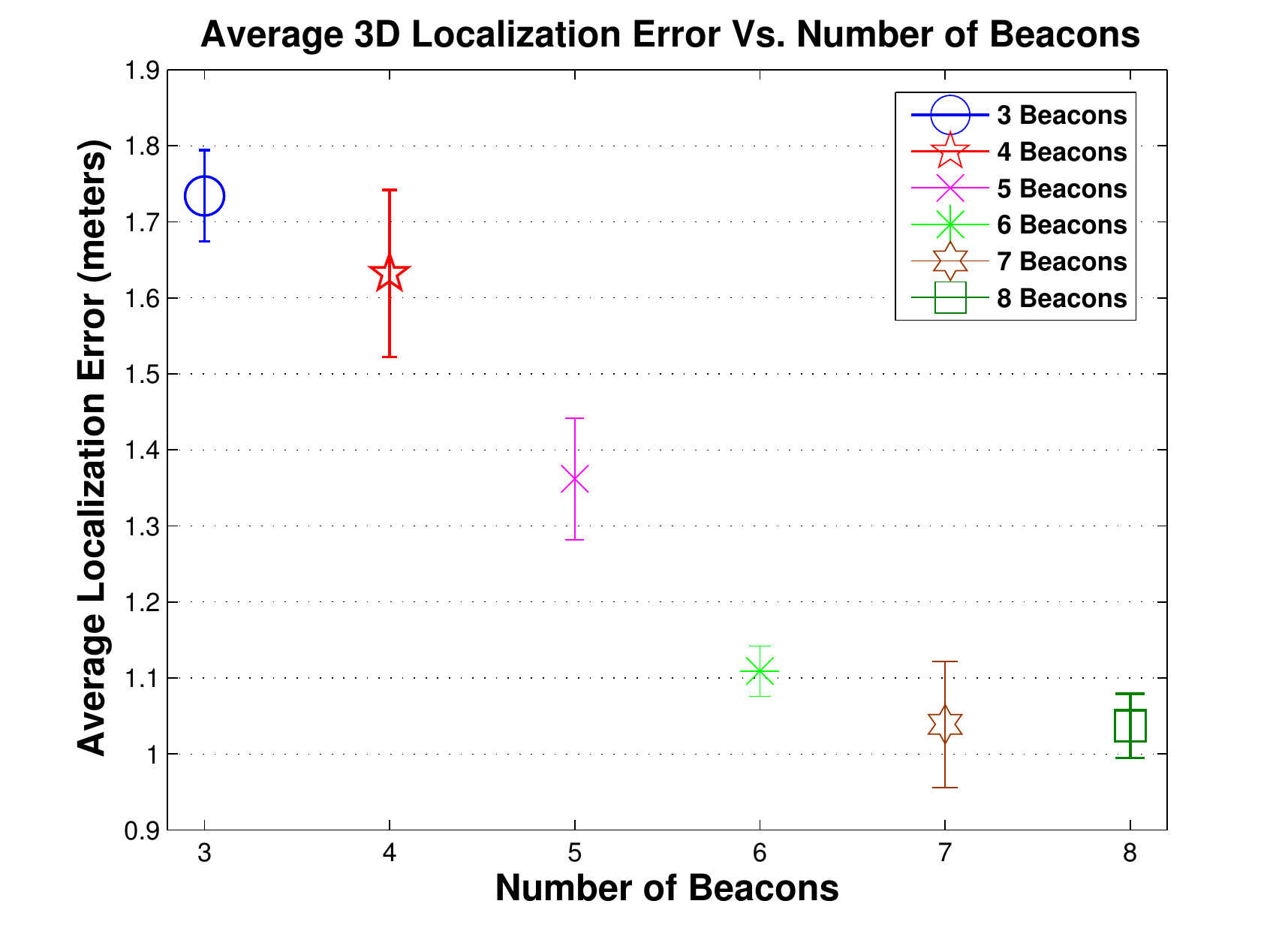}
\caption{Average 3D Localization Error $E_{3D}$ Vs Number of Particles for a varying Number of Beacons for Kalman filter-Particle filter algorithm. }
\protect\label{fig:3dkfpf}
\end{figure} 

\begin{figure}[h!]
\centering
\includegraphics[width=0.46\textwidth]{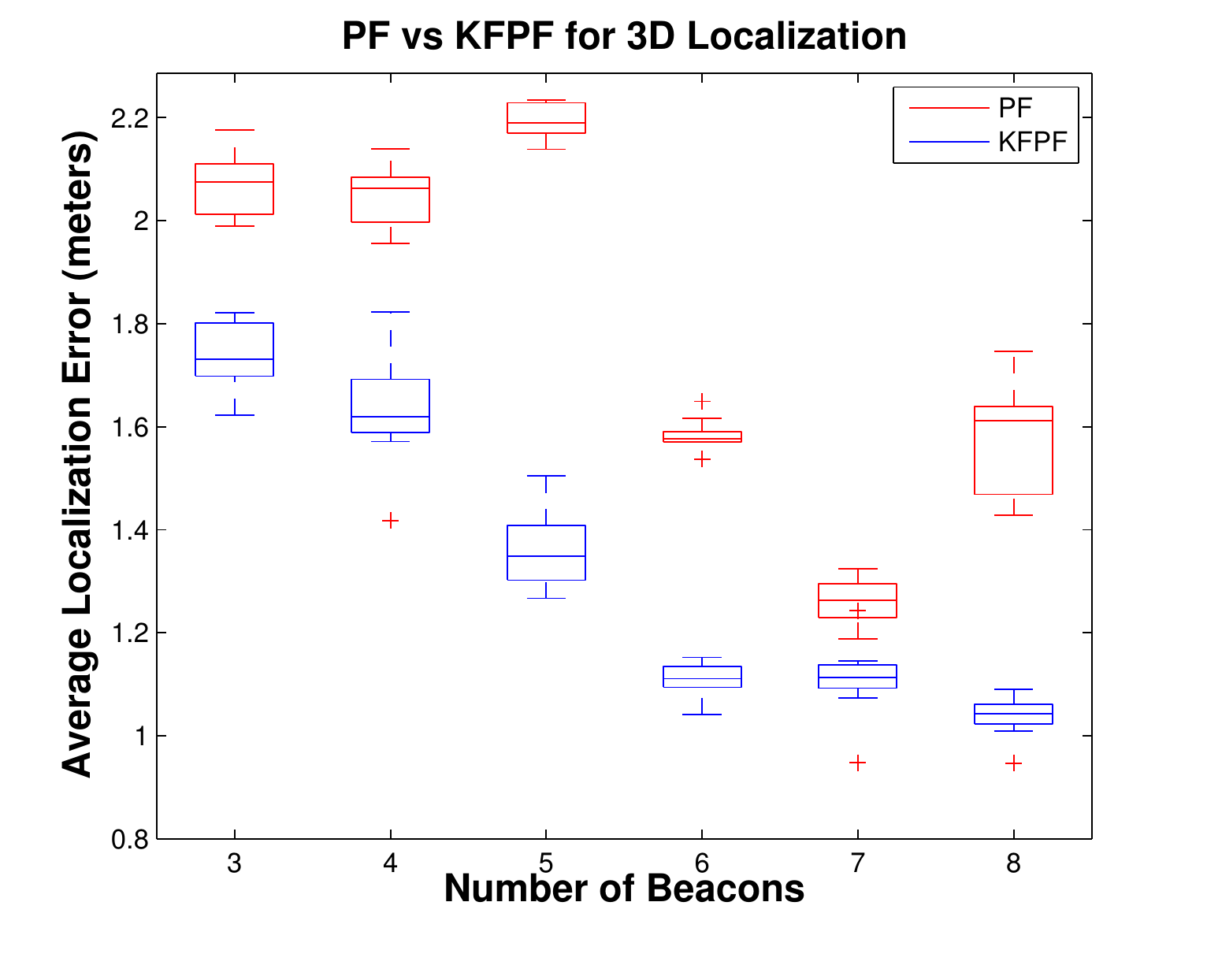}
\caption{Comparison of Particle Filter and Kalman Filter-Particle Filter for 3D localization. }
\protect\label{fig:3dboxplot}
\vspace{-12pt}
\end{figure}

\subsection{3D Localization Results}
Table \ref{tab:pf3d} shows the average 3D localization error $(E_{3D})$ and the standard deviation of $E_{3D}$ for a different number of iBeacons and particles using a Particle filter algorithm. The lowest localization error of 1.189 meters was obtained using seven beacons and 1200 particles. The average 3D localization error using the PF algorithm is 1.78 meters.  Figure \ref{fig:3dpf} shows the  average 3D localization error for a different number of iBeacons. The lowest localization error was obtained using seven iBeacons. However, just as in the case of 2D localization,  the addition of $8^{th}$ iBeacon increased the localization error compared to the seven iBeacon system. In contrast with Figure \ref{fig:pferrorbar}, the results in Figure \ref{fig:3dpf} have a much higher fluctuation due to the $3^{rd}$ \textit{(z)} dimension used in our measurements. 
\par Table \ref{tab:kfpf3d} shows the average 3D localization error $E_{3D}$and the standard deviation of $E_{3D}$ for a different number of iBeacons and particles using our KF-PF cascaded filter based approach. The lowest localization error of 0.94 meters was obtained with 8 iBeacons and 400 particles. The average 3D localization error using KFPF algorithm is 1.33 meters. This is a 25.59\% improvement over the PF algorithm in a 3D environment.  Figure \ref{fig:3dkfpf} shows the average 3D localization error for a different number of iBeacons using a cascaded KF-PF filter. The increase in the number of iBeacons caused a decrease in the average 3D localization error. However, increasing the number of iBeacons from seven to eight did not result in a significant improvement. As mentioned earlier, we believe this is because of the saturation of the experimental space with iBeacons, resulting in self-interference among them. In contrast with Figure \ref{fig:3dpf}, Figure \ref{fig:3dkfpf} shows that the use of cascaded filter reduces the fluctuation in RSSI values and improves the localization accuracy. 
Figure \ref{fig:3dboxplot} compares PF and KFPF  in terms of 3D average localization error for a different number of iBeacons. It can be seen that KFPF outperforms PF in all setups. In Figure \ref{fig:3dboxplot}, the highest improvement is seen with 5 beacons. 

Based on the experimental results that we obtained using our cascaded filter approach and a particle filter, it is evident that cascaded filter based approach results in a lower average localization error, and is a viable approach for indoor localization in both 2D and 3D environments.  The results in Tables \ref{tab:kfpf} and \ref{tab:kfpf3d} in comparison with Tables \ref{tab:pf} and \ref{tab:pf3d} respectively, show that the cascaded filters achieve lower localization error. Furthermore, comparing the results of Figures \ref{fig:boxplot} and \ref{fig:3dboxplot} show that the proposed cascaded filter approach outperforms using only a PF in both 2D and 3D localization for varying number of iBeacons.  This is, as mentioned earlier, because the use of a KF reduced the fluctuation in the RSSI values, which is a major obstacle to localization accuracy. The use of filtered RSSI values with a PF enhanced its localization accuracy and reduced the localization error. Using a cascaded filter in 2D environments, we achieved a localization error as low as 0.708 meters, which is less than the lowest localization error of 0.859 meters achieved using only a PF algorithm. Similarly in a 3D environment, we obtained the lowest localization error of 0.947 meters using our cascaded filter, which is less than the lowest localization error of 1.189 meters achieved with using only a PF algorithm. The use of cascaded filter improved the average 2D localization accuracy by 28\%,  and average 3D localization accuracy by 25.59\% when compared with using only a PF. These results highlight the improvement that is possible when using cascaded filtering approach, and demonstrates that using filtered RSSI values with a PF (our cascaded filter approach) can improve the localization accuracy when compared with using unfiltered RSSI values with a PF.  Also, the algorithmic complexity of PF does not increase significantly when preceded by KF, therefore it is feasible for real-time localization.

\section{Conclusion}
 \label{sec:conclusion}
Indoor localization and proximity can be leveraged to provide a wide range of services including location aware targeted marketing, and indoor navigation. However, the existing solutions proposed in the literature do not fulfil the accuracy, energy efficiency, cost, wide reception range, scalability and availability requirements of both PBS and LBS. In this paper, we proposed an iBeacon based indoor proximity and indoor localization system that has lower energy consumption and higher accuracy. We proposed two algorithms, SRA and SKF, for improving the proximity detection accuracy of iBeacons.  Our experimental results showed that our proposed algorithms SRA and SKF improves the proximity detection accuracy by 29\% and 32\% when compared with the exiting moving average algorithm used by Apple's CoreLocation framework. To leverage iBeacons for indoor localization, we used our KFPF cascaded algorithm to improve the overall localization accuracy when compared with using only PF. KFPF cascaded algorithm improved the average localization accuracy by 28.16\% and 25.59\% in a 2D and 3D environment respectively when compared with using only PF algorithm for localization. 

%

\ifCLASSOPTIONcaptionsoff
  \newpage
\fi

\bibliographystyle{ieeetran}
\bibliography{references}

\end{document}